\documentclass[a4paper,fleqn,usenatbib,useAMS]{mnras}

\usepackage{graphicx}	
\usepackage{amsmath}	
\usepackage{amssymb}	
\usepackage{bm}		

\newcommand{\kms}{\,km\,s$^{-1}$} 

\usepackage{gensymb} 
\usepackage{soul} 

\newcommand{\teff}{T$_{\rm eff}$}

\newcommand{\meta}{\hbox{[Fe/H]}}

\newcommand{\feh}{{\rm {[Fe/H]}}}
\def\aM{\rm [\alpha/Fe]}
\def\afe{\rm [\alpha/Fe]}

\def\kms{\,{\rm km~s^{-1}}}
\def\kpc{\,{\rm kpc}}
\def\mas{\,{\rm mas}}
\def\yr{\,{\rm yr}}

\def\dex{\,{\rm dex}}
\def\Gyr{\,{\rm Gyr}}
\def\vphi{V_\phi}
\def\vphib{v_\phi}
\def\vphipm{V_{\phi,\mathrm{pm}}}

\def\vlos{V_{\rm los}}
\def\vgal{V_{\rm Gal}}
\def\vlsr{V_{\rm LSR}}

\def\ie{{i.e.}\,}
\def\ltsima{$\; \buildrel < \over \sim \;$}
\def\simlt{\lower.5ex\hbox{\ltsima}}
\def\gtsima{$\; \buildrel > \over \sim \;$}
\def\simgt{\lower.5ex\hbox{\gtsima}}

\def\ltsima{$\; \buildrel < \over \sim \;$}
\def\simlt{\lower.5ex\hbox{\ltsima}}
\def\gtsima{$\; \buildrel > \over \sim \;$}
\def\simgt{\lower.5ex\hbox{\gtsima}}

\usepackage[T1]{fontenc}
\usepackage{ae,aecompl}
\usepackage{times}

\title[Rotation fields of the APOGEE survey]
  {Cardinal kinematics: I.  \\Rotation fields of the APOGEE Survey}

\author[G.~Kordopatis et al.]
{Georges~Kordopatis,$^{1}$
Rosemary F.G.~Wyse,$^{2,3}$
Cristina~Chiappini,$^{1}$
\newauthor
 Ivan~Minchev,$^{1}$
 Friedrich~Anders,$^{1}$
 Basilio~Santiago$^{4,5}$
\\
$^{1}$Leibniz-Institut f\"ur  Astrophysik Potsdam (AIP), An der Sternwarte 16, 14482 Potsdam, Germany\\
$^{2}$Johns Hopkins University, Department of Physics \& Astronomy, 3400 N Charles Street, Baltimore, MD 21218, USA\\
$^{3}$University of Edinburgh, Institute for Astronomy, Edinburgh, EH9 3HJ, UK\\
$^4$Instituto de F\'isica, Universidade Federal do Rio Grande do Sul, Caixa Postal 15051, 91501-970 Porto Alegre, Brazil\\
$^5$Laborat\'orio Interinstitucional de e-Astronomia - LineA, Rua Gal. Jos\'e Cristino 77, 20921-400 Rio de Janeiro, Brazil\\
}

\pagerange{\pageref{firstpage}--\pageref{lastpage}} \pubyear{2015}

\def\LaTeX{L\kern-.36em\raise.3ex\hbox{a}\kern-.15em
    T\kern-.1667em\lower.7ex\hbox{E}\kern-.125emX}

\begin{document}

\label{firstpage}

\maketitle

\begin{abstract}
Correlations between stellar chemistry and kinematics have
long been used to gain insight into the evolution of the Milky Way
Galaxy.  Orbital angular momentum is a key physical parameter and it is often estimated from three-dimensional space motions. We here demonstrate the lower uncertainties that can be
achieved in the estimation of one component of  velocity through selection of stars in key directions and use of line-of-sight velocity alone (i.e. without incorporation of  proper motion data). In this first paper we apply our technique to stars observed in the
direction of Galactic rotation in the APOGEE survey. We first derive the distribution of azimuthal velocities, $\vphib$, then from these and observed radial coordinates, estimate the stellar guiding centre radii, $R_g$, 
within  $6.9\leq R \leq 10\kpc$  with uncertainties smaller than (or of the order of) $1\kpc$.
We show that there is no simple way to select a clean stellar sample based on low errors on proper motions and distances to obtain high-quality 3D velocities and hence  one should pay particular attention when trying to identify kinematically peculiar stars based on velocities derived using the proper motions. Using our $\vphib$ estimations, we   investigate the joint distribution of elemental abundances and rotational kinematics free from the blurring effects of epicyclic motions, and we derive the $\partial \vphib / \partial \afe$ \& $\partial \vphib / \partial \feh$  trends for the thin and thick discs as a function of radius. Our analysis   provides further evidence for radial migration within the thin disc and hints against radial migration playing a significant role in the evolution of the thick disc.
 
\end{abstract}

\begin{keywords}
Galaxy: abundances -- Galaxy: disc -- Galaxy: kinematics and dynamics -- Galaxy:
stellar content -- Galaxy: evolution.
\end{keywords}

\section{Introduction}
Correlations between stellar chemistry and
kinematics, as well as the way these quantities change across
the Galaxy, provide valuable insight into  how the Milky Way
formed and evolved \citep[e.g.][]{Eggen62,Freeman02}. Full six-dimensional (6D)
kinematic-position phase space information is obviously preferred for
such analyses; this requires proper motions and distances, in addition
to line-of-sight velocities.  In the pre-Gaia era, for stars
beyond the immediate solar neighbourhood distances must in
general be obtained through isochrone fitting
 \citep[e.g.][]{Pont04, Jorgensen05, Binney14a} and only ground-based
measures of proper motions are available. The
uncertainties in the estimated values of each of these quantities
are relatively high. Indeed ground-based proper motion catalogues such as PPMXL
or UCAC4 \citep{Roeser10,Zacharias13} have random  errors of the order of
$4-10 \mas\yr^{-1}$, resulting in transverse velocity errors
of $60-150\kms$ for a star with a perfectly known
distance of $3\kpc$ from the Sun\footnote{Gaia will provide at least an order-of-magnitude improvement in proper-motion errors.}. 
Typical uncertainties in
distances obtained through isochrone-fitting are 15-30~percent,
further increasing the error in transverse velocity. In comparison, line-of-sight
velocities can routinely be obtained with uncertainties of below
$1\kms$, so that the transverse velocity dominates the error budget in
the derived 3D space motion.


The higher precision and accuracy of line-of-sight velocities motivate spectroscopic studies of stars
in selected key directions  where the line-of-sight velocities are
particularly sensitive to one component of space motion.  These include lines-of-sight at low-to-moderate Galactic latitudes 
towards the Galactic centre and anti-centre (Galactic longitudes $\ell
\sim 0^\circ, \, 180^\circ$ respectively) which probe velocities along the radial coordinate; towards the Galactic Poles (latitudes $|b| \sim 90^\circ$), which probe velocities perpendicular to the plane;  
and low-to-moderate Galactic latitudes towards and against Galactic rotation (Galactic longitudes $\ell
\sim 90^\circ; \; 270^\circ$ respectively) which probe azimuthal velocities.

In this paper, the first of a series, we analyse the joint
distributions of azimuthal velocity and elemental abundances of stars
in the \lq rotation' fields of the Apache Point Observatory Galactic Evolution Experiment \citep[APOGEE,][]{Majewski15}. In subsequent papers of this  series we will apply the techniques developed below  to other large spectroscopic surveys, namely RAVE \citep{Steinmetz06} and Gaia-ESO \citep{Gilmore12}, and extend the analysis to the other cardinal directions. The azimuthal velocity is of course the component with the largest expected mean value, measuring the orbital angular momentum.  Angular momentum
about the $Z-$axis is an integral of motion in an axisymmetric system
and is often still a meaningful quantity in more realistic
potentials. The surface elemental abundances are conserved through most of the lifetime of most low-mass stars (excepting those in close binary systems with mass transfer) and reflect the abundances in the gas from which the star formed, and hence constrain its birthplace. The combination of azimuthal velocity and elemental abundances provides insight into dynamical processes such as the creation of moving groups - whether as tidal debris from a disrupted
satellite/star cluster or created through resonant interactions with
gravitational perturbations  - and radial migration of stars through the disc \citep[e.g.][]{Dehnen98,Helmi99,Sellwood02,Antoja08,Minchev10b}.

\bigskip

The paper is structured as follows:  in Sect.~\ref{sec:method} we present the method that we will use to derive reliable measures of the  azimuthal velocities without the use of proper motions and we test it by means of  a mock catalogue derived from the Besan\c con model \citep{Robin03}.  In Sect.~\ref{sec:application_to_data} we apply this method to the APOGEE-Data Release 12 catalogue \citep{Alam15,Holtzman15}. Section~\ref{sect:chemodynamics} presents our investigation of the relationships between the derived kinematics and chemical abundances for this sample and in Sect.~\ref{sect:thick} and Sect.~\ref{sect:constraints_churning} we interpret these results for the thick disc and thin disc, respectively. Our conclusions are given in Sect.~\ref{sect:conclusions}.

\section{Estimator of azimuthal velocity}
\label{sec:method}
We obtain the estimator of the azimuthal velocity from the line-of-sight velocity alone following  the derivation presented in \citet[][Sect.\,5]{Morrison90}, based upon \cite{FW80} \citep[see also][and more recently \citealt{Kordopatis13a}]{Wyse06}. For convenience, we here repeat the main steps from  \cite{Morrison90}, with our equations (1-10) below being essentially their equations (1-8) (plus discussion).

\subsection{Estimator of azimuthal velocity for a  given star}
\label{sec:mean_vphi}

The Galactocentric line-of-sight velocity of a star observed with coordinates $(\ell, b)$, denoted by $\vgal$, is obtained from  its heliocentric line-of-sight velocity ($\vlos$) by correcting for the velocity of  the local standard of rest ($\vlsr$) and for the Sun's peculiar motion projected into that line-of-sight ($v_{{\rm pec},\odot,(\ell,b)}$). Therefore we have: 
\begin{equation}
\vgal=\vlos +  v_{{\rm pec},\odot,(\ell,b)} + \vlsr\sin \ell \cos b.
    \end{equation}
 We adopt  $\vlsr=220\kms$ \citep{Kerr86} and $v_{{\rm pec},\odot,(\ell,b)}$ is defined as:
    \begin{equation}
  v_{{\rm pec},\odot,(\ell,b)} = U_0\cos \ell \cos b + V_0\sin \ell \cos b +W_0 \sin b,
    \end{equation}
    where the values $(U_0,V_0,W_0)=(11.1,12.2,7.2)\kms$ are taken from \citet{Schonrich10}.

    In spherical coordinates $(r,\phi,\theta)$, $\vgal$ may be written in terms of the true components of space motion as:
    \begin{equation}
    \vgal=\alpha V_r +\beta V_\phi + \gamma V_\theta.
    \label{eq:Vgal_sph}
    \end{equation}
     The values of the coefficients ($\alpha,\, \beta,\, \gamma$) depend on the star's position in the Galaxy, and can be expressed as a function of the Galactic coordinates $(\ell,b)$,  the line-of-sight distance of the star relative to the Sun ($d$), and the Sun's distance to the Galactic centre ($R_0=8\kpc$) as: 
     
\begin{eqnarray}
\alpha&=& \frac{1}{r} (d-R_0\cos b \cos \ell) \\
\beta &=& \frac{1}{R} R_0 \cos b \sin \ell \label{eq:beta_definition} \\
\gamma &=& \frac{\sin b}{rR}(dR_0\cos b \cos \ell -R_0^2)
\end{eqnarray}
where $r$ is the Galactocentric distance of the star and $R$ is the projection of this distance on the Galactic plane:  
     \begin{eqnarray}
r^2 &=& R_0^2+d^2 - 2dR_0\cos b \cos \ell \\
R^2 &=& R_0^2+d^2 \cos^2 b - 2dR_0 \cos b \cos \ell.
 \label{eq:u2}
       \end{eqnarray} 
 Assuming that the mean stellar motions in $r$ and $\theta$ can be approximated to be zero (i.e the Galaxy is in equilibrium), then the quantity $\vphib$, defined as
     
\begin{equation}
\vphib=\vgal/\beta
\label{eq:vphib}
\end{equation}
is  an unbiased estimator of the true azimuthal velocity $V_\phi$ of the star.

\begin{figure}
\centering
\includegraphics[width=0.9\linewidth, angle=0]{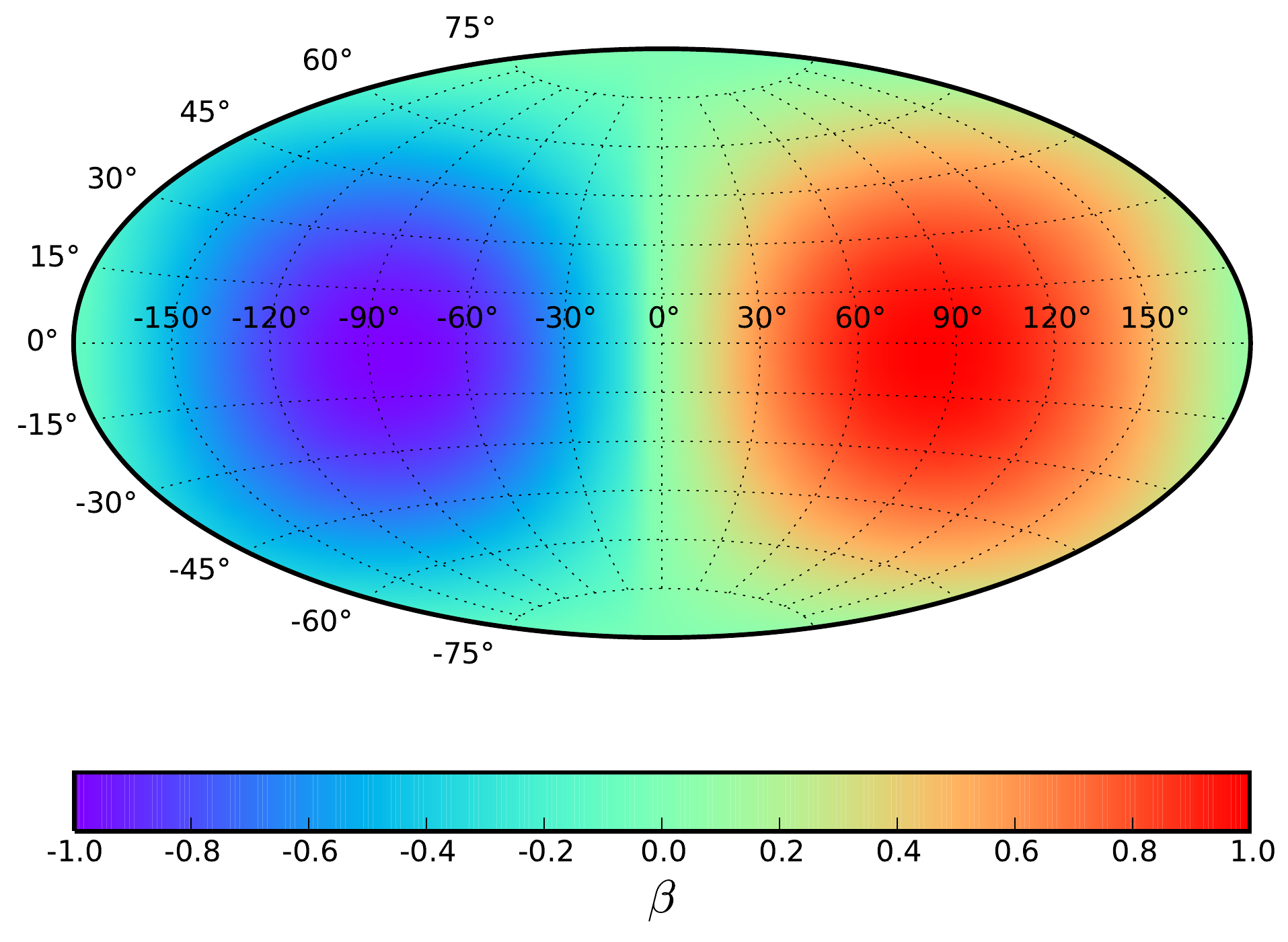}
\caption{Aitoff projection in Galactic coordinates of the variation of the $\beta$ parameter with line-of-sight, illustrating the changing contribution of $\vphi$ to $\vlos$ for stars at $1\kpc$. }
\label{fig:Aitoff_beta}
\end{figure}

\begin{figure}
\centering
\includegraphics[width=0.95\linewidth, angle=0]{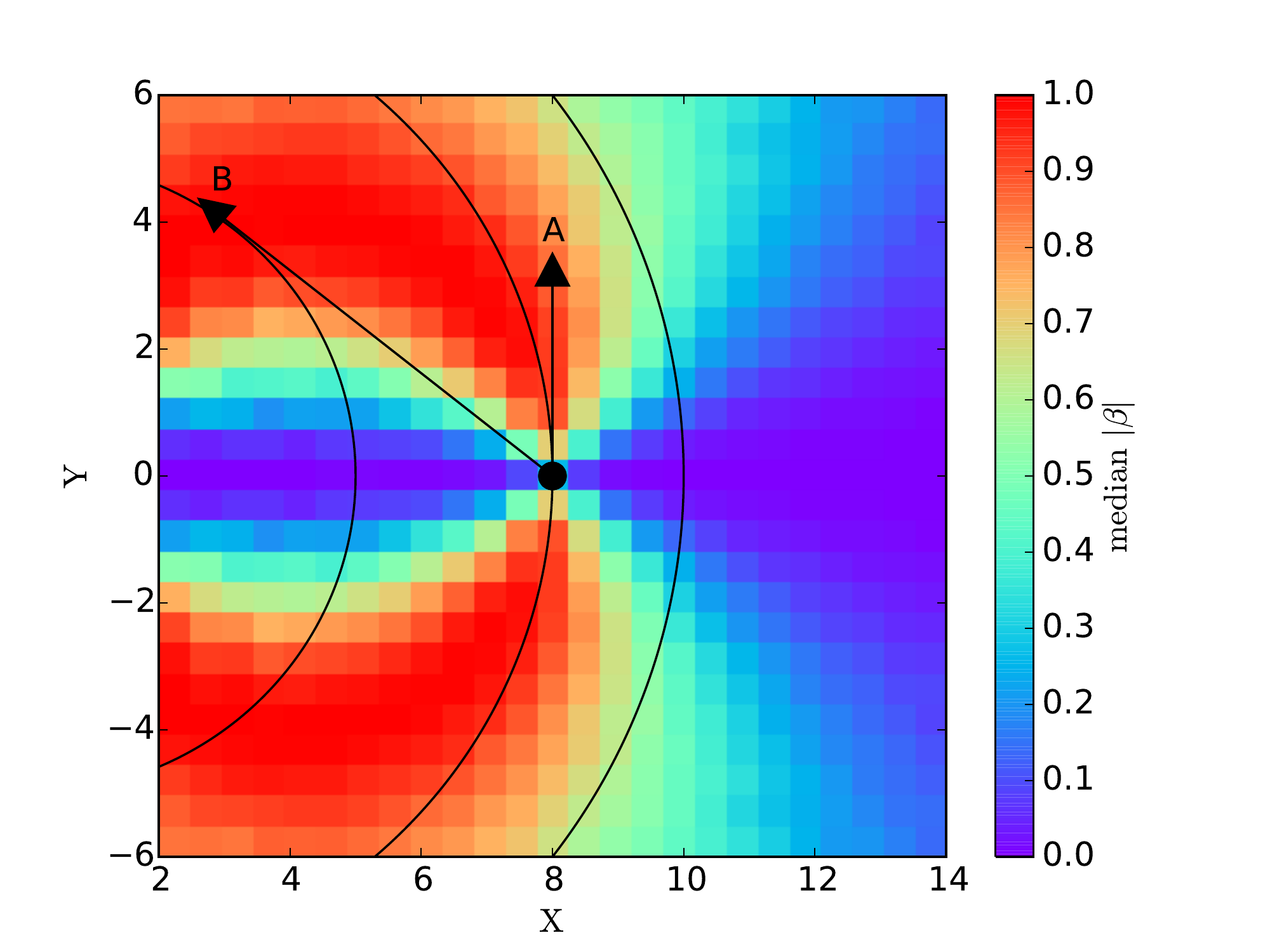}
\caption{ $|\beta|$ variations within the Galactic plane (cartesian coordinates $X-Y$). The Sun is located at $(X,Y)=(8,0)\kpc$. Three circular orbits with a radius of $R=5, 8, 10\kpc$ are drawn, in order to illustrate how the line-of-sight velocity probes the true azimuthal velocity for stars on these orbits. Line-of-sight `A' probes $\vphi$ through $\vlos$ for a continuous range of circular orbits and is not very sensitive to distance errors, whereas line-of-sight `B' probes only small radii with high $\beta$ and is more sensitive to distance errors. }
\label{fig:XY_beta}
\end{figure}

It is clear that for $\beta \sim 1$ the line-of-sight velocity 
is maximally sensitive to the true azimuthal velocity $\vphi$ (or, in other words, that there is no contamination from the other velocity components other than $\vphi$).  
Figure~\ref{fig:Aitoff_beta} illustrates the variation of $\beta$ as a
function of the Galactic sky coordinates $(\ell,b)$, for an assumed
distance (from the Sun) of $d=1\kpc$. Complementary to
Fig.~\ref{fig:Aitoff_beta}, Fig.~\ref{fig:XY_beta} shows the
variations of $|\beta|$ for stars in the Galactic plane, at various
distances from the Sun: towards the inner Galaxy there are several
lines-of-sight for which $\vlos$ is an excellent probe of  the azimuthal
velocity of the stars. However, as we will see in
Sect.~\ref{sec:Besancon}, not all of the Galactic longitudes have the
same sensitivity to either distance errors (random and systematic) or
deviations from true Solar position $R_0$. Furthermore, it is only
towards the directions of Galactic rotation ($\ell=90^\circ$) and
anti-rotation ($\ell=270^\circ$) that $\vphib$ can be obtained for a
continuous range of radial distance, i.e. without crossing regions of
the Galaxy where $\vphib$ is not a good approximation for $\vphi$
(e.g., line-of-sight `B' in Fig.~\ref{fig:XY_beta}).

\subsection{Estimator of mean azimuthal velocity}

The parameter $\vphib$ (defined above as $V_{Gal}/\beta$) is the unbiased estimator
of the true azimuthal velocity $V_\phi$ for any single star. However,
as discussed in \citet{FW80} and in \citet{Morrison90}, the unbiased mean
azimuthal velocity of a population of stars should be 
estimated using  the
quantity $\beta V_{Gal}$, which gives highest weight to the stars in the lines-of-sight 
with highest sensitivity to the true $V_\phi$ (highest values of $| \beta |$), rather than by taking the mean of all the individual values of $\vphib$. Thus for a sample of stars, each with projection factor $\beta_i$ and Galactocentric line-of-sight velocity $V_{Gal,i}$, the estimator of the mean azimuthal (rotation) velocity, denoted by $\overline{V_{rot}}$, is:

\begin{equation}
\overline{V_{rot}} = \frac{\sum_i\beta_i V_{Gal,i}}{\sum_i \beta_i^2}
\label{eq:vrot}
\end{equation}
where the sum is over all the stars in the sample. 
Pruning of outliers can be performed to increase the robustness of the derived mean \citep[see][]{Morrison90}, but the location of the cut is rather subjective and we have chosen to use the full sample.


  \subsection{Tests using the Besan\c con model}
  \label{sec:Besancon}
  We generated a synthetic data set using the Besan\c con model\footnote{The Besan\c con model assumes the following parameter values: $\vlsr=226.4\kms$, $(U_0,V_0,W_0)=(10.3,\, 6.3,\, 5.9)\kms$  and $R_0=8.5\kpc$. In order to be self-consistent, in this section we use these values for these parameters,  rather than those currently accepted values that we adopted in our analysis of the observational data below.}
of the Galaxy \citep{Robin03}. The model provides proper motions,
heliocentric line-of-sight velocities ($\vlos$), distances
and each of the  components of the 3-dimensional velocities of the generated stars. This represents
an ideal case with which to test how well our estimator recovers  the true
azimuthal velocities.

  \subsubsection{Global trends ($10^\circ \leq \ell \leq 350^\circ$)}
\label{subsub:trends_all_l}

 \begin{figure}
\centering
\includegraphics[width=0.9\linewidth, angle=0]{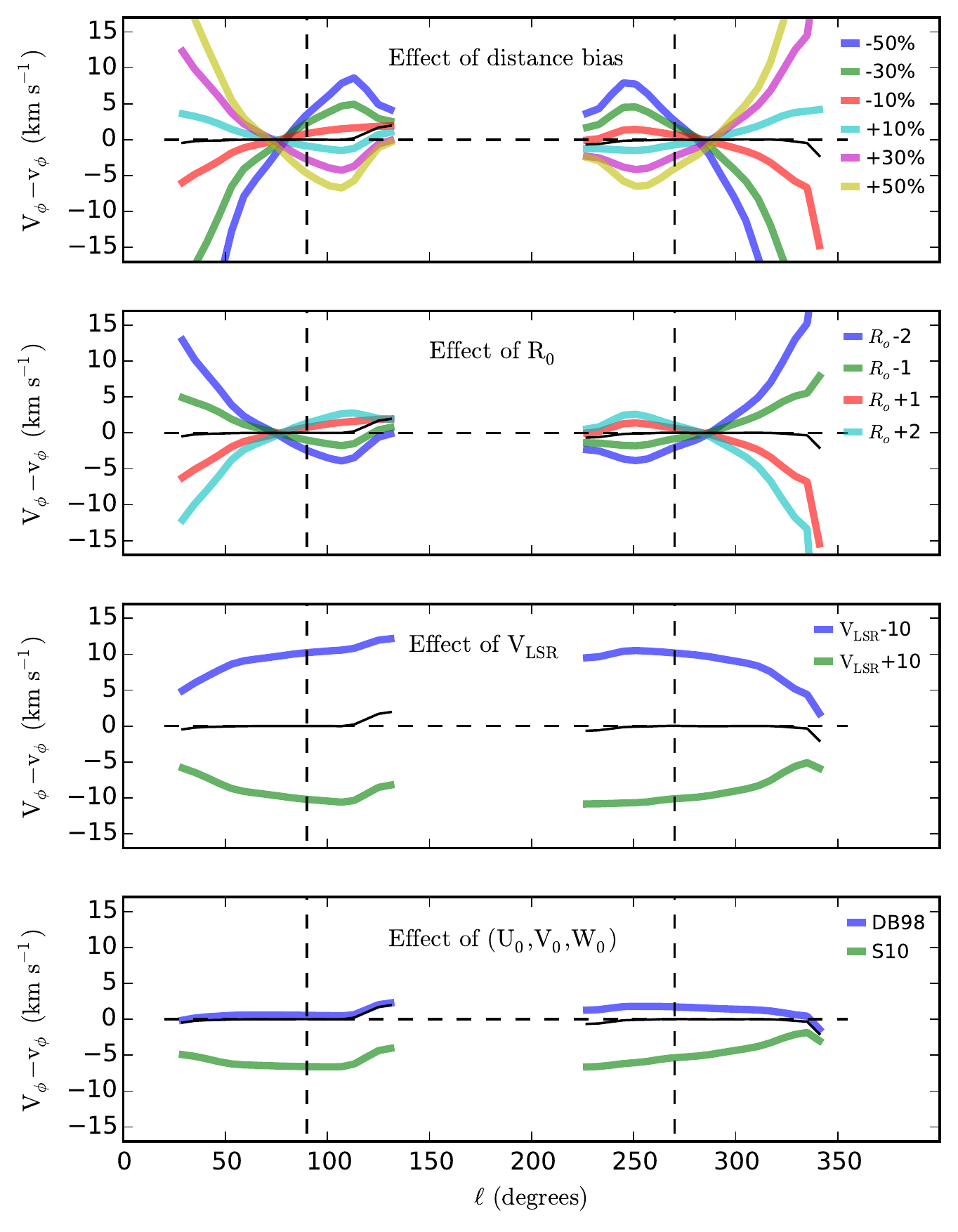}
\caption{  Sensitivity of the $\vphib$ estimator to the accuracy of the stellar distances (top row), solar location $R_0$ (second row), local standard of rest velocity (third row) and Solar motion $(U_0, V_0, W_0)$ (bottom row), all as a function of Galactic longitude. The tests are done using simulated stars from the Besan\c con model (sample spanning all $\ell$ and $b$) having $\beta\geq0.75$. In each panel, the input parameters from the model are considered to be the `true' parameters.   Thin black curves are the estimations for perfectly well known distances, location of the Sun, $\vlsr$ and $(U_0, V_0, W_0)$.  }
\label{fig:Besancon_vphi_sensitivity_all_glon}
\end{figure}

We have selected mock catalogues towards $10^\circ \leq \ell \leq 350^\circ $ and $15^\circ  \leq b \leq 45^\circ $ (with $10^\circ$ steps in $\ell$ and $b$). The range in $b$ has been chosen in order to match the intermediate latitudes targeted in optical surveys, albeit that  the contribution of $\vphi$ to $\vlos$ is smaller than for latitudes closer to the Galactic plane.
We then estimated the difference between the $\vphib$ estimator and the true $\vphi$ as a function of Galactic longitude (Fig.~\ref{fig:Besancon_vphi_sensitivity_all_glon}). This was done for stars  having $|\beta|>0.75$, firstly assuming perfectly known distances and $\vlos$   (thin black lines in all subplots of Fig.~\ref{fig:Besancon_vphi_sensitivity_all_glon}), and secondly for biased distances (up to $\pm50$ per cent, top row), incorrect assumed value for the Solar location $R_0$ ($\pm2\kpc$, second row), $\vlsr$ ($\pm10\kms$, third row), and  Solar motion $(U_0,V_0,W_0)$ (adopting the values of either \citealt{Dehnen98} or \citealt{Schonrich10}, bottom row).  
We discuss the outcome of each of these points below. 

\smallskip

(i) For perfectly known distances and $\vlos$, all the lines-of-sight provide unbiased estimations of true $\vphi$, as it can be seen from the thin black line in each plot.

(ii) 
For lines-of-sight close to the cardinal directions ($\ell=\pm90^\circ$) distance biases and deviations from true $R_0$ do not introduce global offsets from $\vphi$ (see Sect.~\ref{subsub:cardinal_direction} for further details). Towards lines-of-sight that differ more than $\sim \pm40^\circ$ from the direction of Galactic rotation and anti-rotation (indicated by vertical dashed lines in Fig.~\ref{fig:Besancon_vphi_sensitivity_all_glon}), distance biases as small as 10 per cent already introduce offsets from true $\vphi$ of the order of $\sim 5\kms$. These offsets do not have the same sign depending on whether the los is before or after the cardinal directions and depending on whether it probes Galactic rotation or anti-rotation. 
This strong $\ell$ dependence can be better understood by having a closer look at Fig.~\ref{fig:XY_beta}. 
 Towards the inner Galaxy, biased distances  lead to an over/under-estimation of the $|\beta|$ values. This implies that stars that should not be considered in the computation of $\vphib$ due to their location are erroneously placed at radii  that probe  $\vphi$ well, and vice versa.

(iii) Adopting an incorrect value of $\vlsr$ (by $\pm10\kms$) will globally alter in a similar way the estimation of $\vphi$. In that case,  the offsets  have a smaller dependence on $\ell$ compared to distance biases or erroneous value of $R_0$. 

(iv) The same conclusion as for point (iii) can be drawn for incorrect assumed values for  $(U_0, V_0, W_0)$. In particular, offsets in $V_0$ will have a direct impact on the derived $\vphib$, but the $\ell$ dependence is small, below $5\kms$.   

\smallskip
Given the points above, for the remainder of the paper we will therefore focus towards the cardinal directions of Galactic rotation and anti-rotation.

  \subsubsection{Trends towards $\ell\sim90^\circ$ and $\ell\sim 270^\circ$}
  \label{subsub:cardinal_direction}

  \begin{figure}
\centering
\includegraphics[width=0.9\linewidth, angle=0]{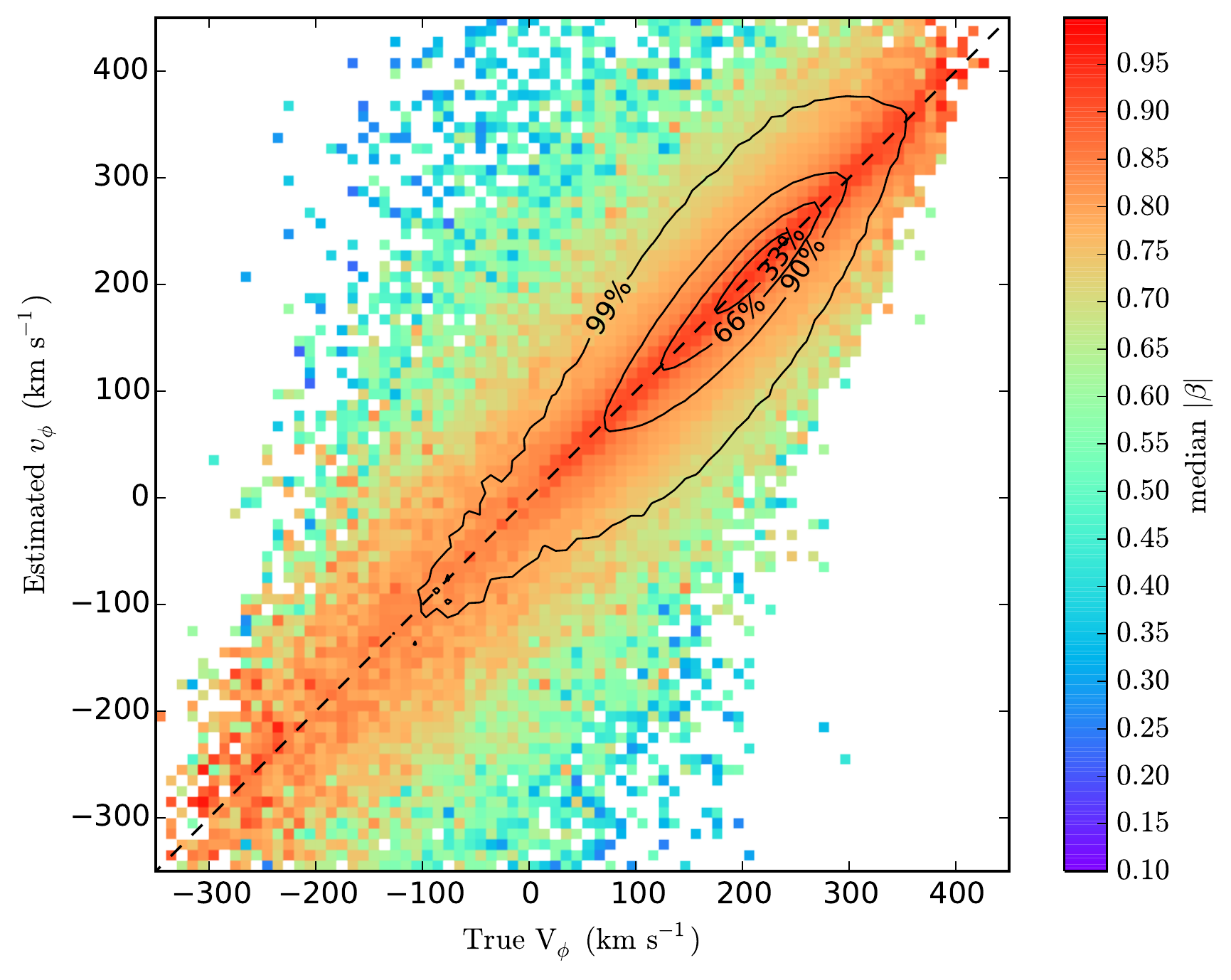}
\caption{Comparison of the $\vphib$ estimator with the true azimuthal velocity $\vphi$ obtained from the Besan\c con model with the simulated stellar sample selected to lie between $245^{\circ} \leq \ell \leq 295^{\circ}$ and $13^{\circ}\leq b \leq 43^{\circ}$,   spanning distances up to $50\kpc$ and hence including low values of $|\beta|$. The colour-code is the median of the $|\beta|$ parameter in a given $( \vphi , \; \vphib)$ bin.  One  can see that stars with  lower values of $|\beta|$ deviate the most from the 1:1 relation. Distances and $\vlos$ are assumed to be perfectly known here. Table~\ref{tab:errors_vphi} quantifies the means and dispersions of the distribution of the offset $(\vphi-\vphib)$ as a function of $\beta$.}
\label{fig:Besancon_vphi}
\end{figure}

Figure~\ref{fig:Besancon_vphi} shows the results obtained for a mock sample covering a sky region defined
within  $245^{\circ} \leq \ell \leq 295^{\circ}$ and $15^{\circ}\leq b \leq 45^{\circ}$, assuming perfectly known distances and $\vlos$.  
 The range in $\ell$ for the mock sample of Fig.~\ref{fig:Besancon_vphi} has been selected in order to cover  $\pm25^\circ$ against the direction of Galactic rotation (similar results are obtained for longitudes towards  the direction of Galactic rotation,  or for negative latitudes). The simulated  stellar sample generated with this selection function spans line-of-sight distances up to $50\kpc$  (mean distance $2.3\kpc$ and 80 per cent of the sample closer than $3.2\kpc$), and for this range of $(\ell,b)$ and distances, the value of $\beta$ ranges from $0.14$ to $0.97$  \citep[the low $\beta$ values mostly reflecting the larger distances, see][]{Fermani13}. 

An immediate confirmation from Fig.~\ref{fig:Besancon_vphi}, quantified in Table\,\ref{tab:errors_vphi}, is
that the estimator $\vphib$ (Eq.\,\ref{eq:vphib}) does indeed provide an unbiased estimate of the
true azimuthal velocity,  even for $|\beta|$  as low as $\sim0.6$. 
The scatter about the relation
between the true and estimated values of the velocity clearly decreases with increasing $|\beta|$ (again quantified in
Table~\ref{tab:errors_vphi} for different ranges of
$|\beta|$), reaching a minimum value of $10\kms$ for $|\beta|>0.95$.

\begin{table}
 \caption{Bias and dispersion in azimuthal velocity estimations as a function~of~$|\beta|$ for stars with perfectly known distances and $\vlos$.}
\begin{center}
\begin{tabular}{cccc}\hline \hline
$|\beta|$-range & ${\vphi-\vphib}$ & $\sigma_{(\vphi-\vphib)} $ \\
			& $\kms$ & $\kms$ \\ \hline		
$[0.60,0.65]$ & $0.7$ & $75$ \\
$[0.65, 0.70]$ & $-0.2$ & $56$ \\
$[0.70,0.75]$& $-0.1$  & 45 \\
$[0.75,0.80]$& $0.0$ & 36 \\
$[0.80,0.85]$& $-0.0$ & 29 \\
$[0.85,0.90]$& $-0.1$ & 25 \\
$[0.90,0.95]$& $0.0$ & 18 \\
$[0.95,1]$     & $0.0$ & 10 \\
\hline
\end{tabular}
\end{center}
\label{tab:errors_vphi}
\end{table}%

As shown in Table~\ref{tab:errors_vphi_realistic}, these offsets and dispersions are not altered when introducing realistic uncertainties on distances obtained from isochrone fitting ($\sim$30 per cent), nor when introducing uncertainties on the $\vlos$ up to $10\kms$ (corresponding to the  lowest $\sigma_{(\vphi-\vphib)} $ value reached for perfectly well known distances and $\vlos$, see Table~\ref{tab:errors_vphi}). Such uncertainties are easily reached by most of the major large stellar spectroscopic surveys, including those with low \citep[SEGUE,][]{Yanny09} and medium \citep[RAVE,][]{Kordopatis13b} resolution.
 In the analysis below we will hence adopt the dispersion for a given range of
$| \beta |$ as the amplitude of the typical error on $\vphi$ for a
single star with $| \beta |$ in that range. Further, the results shown in
Table~\ref{tab:errors_vphi_realistic} were obtained using  generic selection criteria and can
therefore be used to set a threshold minimum acceptable value of 
$|\beta|$. This threshold can be tuned for each survey, based on the
trade-off between increasing the number of selected targets and decreasing the uncertainty in  the estimated rotation velocity.

\begin{table}
 \caption{Bias and dispersion in azimuthal velocity estimations as a function~of~$|\beta|$ for stars with random errors in distances of 30 per cent and $\vlos$ errors of $0.2\kms$ and $10\kms$.}
\begin{center}
\begin{tabular}{c|cc|cc}\hline \hline
 &\multicolumn{2}{c}{$\sigma_{\vlos} = 0.2\kms$} &\multicolumn{2}{c}{$\sigma_{\vlos} = 10\kms$} \\
$|\beta|$-range & ${\vphi-\vphib}$ & $\sigma_{(\vphi-\vphib)} $ & ${\vphi-\vphib}$ & $\sigma_{(\vphi-\vphib)} $ \\
			& $\kms$ & $\kms$ & $\kms$ & $\kms$ \\ \hline		
$[0.60,0.65]$ & $0.3$ & $74$ &$-8.7$& 74 \\
$[0.65, 0.70]$ & $-1.1$ & $57$ &$-5.2$& 58 \\
$[0.70,0.75]$& $-0.3$  & 45 &$-2.5$& 47 \\
$[0.75,0.80]$& $-0.4$ & 36 &$-0.9$& 38 \\
$[0.80,0.85]$& $-0.4$ & 30 &$-0.2$& 32 \\
$[0.85,0.90]$& $-0.6$ & 26  &$-0.6$& 28 \\
$[0.90,0.95]$& $-0.6$ & 19 &$-0.1$& 21 \\
$[0.95,1]$     & $-0.7$ & 11 &$0.5$& 15 \\
\hline
\end{tabular}
\end{center}
\label{tab:errors_vphi_realistic}
\end{table}%

 \begin{figure*}
\centering
\includegraphics[width=0.9\linewidth, angle=0]{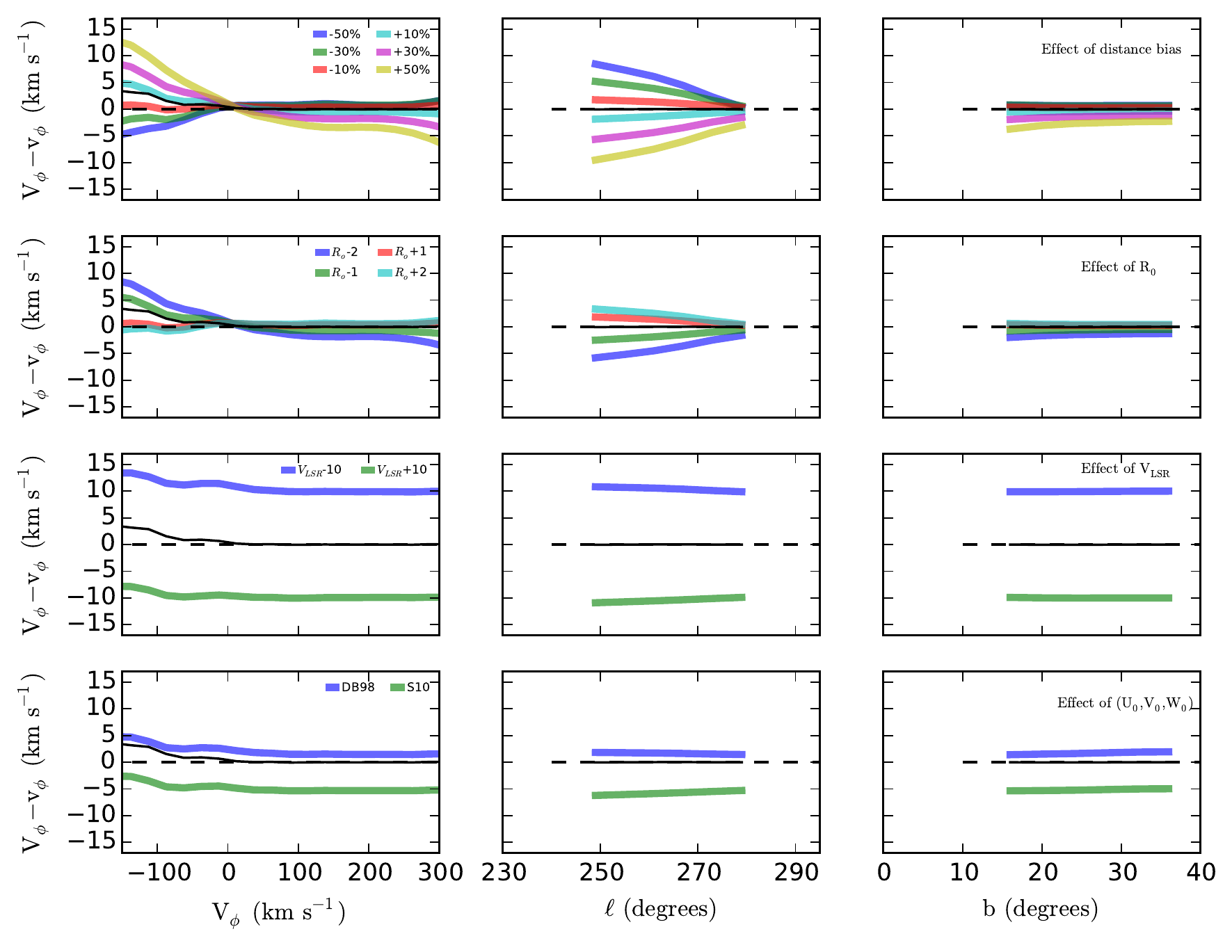}
\caption{Sensitivity of the $\vphib$ estimator to the accuracy of the stellar distances (top row), the Sun's location $R_0$ (second row), local standard of rest velocity (third row) and Solar motion $(U_0, V_0, W_0)$ (bottom row). Tests are done using the Besan\c con simulated stars (same as in Fig.~\ref{fig:Besancon_vphi}) having $\beta\geq0.75$. In each panel, the input parameters from the model are considered to be the `true' parameters.  Each plot shows the mean difference between true $\vphi$ and the estimated one,$\vphib$, as a function of $\vphi$ (left column), Galactic longitude $\ell$ (middle column), and Galactic latitude $b$ (right column). Thin black curves are the estimations for perfectly well known distances, Sun's position, $\vlsr$ and $(U_0, V_0, W_0)$.  }
\label{fig:Besancon_vphi_sensitivity}
\end{figure*}

\bigskip
Figure~\ref{fig:Besancon_vphi_sensitivity} is a more thorough analysis of the $\vphib$ biases towards Galactic anti-rotation. The offsets of our estimator are illustrated as a function of true $\vphi$ (left column), Galactic longitude (middle column) and Galactic latitude (right column). In addition to the comments made in Sect.~\ref{subsub:trends_all_l} for $\vlsr$ and $(U_0,V_0,W_0)$, the following conclusions can be reached:

(i) Over (under) estimations of line-of-sight distances result in an
under (over) estimation of $\vphi$ for counter-rotating stars (those with negative values), and an
over (under) estimation of $\vphi$ for the fast rotators (moving faster than the Sun). This is due
to the fact that $\beta$ decreases (increases) when biasing the
distances and hence increases (decreases) the absolute value of
$\vgal/\beta$ (see Eq.~\ref{eq:beta_definition}).  However, even when
the assumed  biases are as high as 50 per cent, the difference in
$\vphib$ does not,   in most cases, exceed $\sim \pm10\kms$. In
particular, the variation as a function of $\ell$ is smaller than $5\kms$
for distance biases smaller than 30 per cent and for the 
relevant range of Galactic longitudes. Finally, the effect as a function of
Galactic latitude remains below $\sim3\kms$ for $b\leq
45^\circ$. Overall, for distance biases below 10 per cent, all the
effects that are introduced in the $\vphib$ estimator are negligible.

(ii) The same reasoning as for point (i) can be followed for the effect of $R_0$ on the estimated $\vphib$. An over-estimated $R_0$ will result to an increased $|\beta|$ (see Eq.~\ref{eq:beta_definition}) and hence a $\vphib$ value which will be an under-estimation of true  $\vphi$ for high velocity stars and an over-estimation for counter-rotating stars. Nevertheless, here again, even an offset of $1\kpc$ from the true $R_0$ will not result in  effects greater than $5\kms$. 

\smallskip

From the above points one can conclude that our estimator is robust. Indeed, provided that the stellar distances are not biased by more than 10-20 per cent, that $R_0$ is known to better than $1\kpc$ and that the assumed $\vlsr, U_0, V_0, W_0$ are sound, no strong systematics depending on the star's position in the Galaxy are introduced that cannot safely be ignored in our analysis, provided that the  lines-of-sight included remain close to the directions of Galactic rotation and anti-rotation.

\subsubsection{Kinematics-metallicity relations}

\begin{figure}
\centering
\includegraphics[width=0.85\linewidth, angle=0]{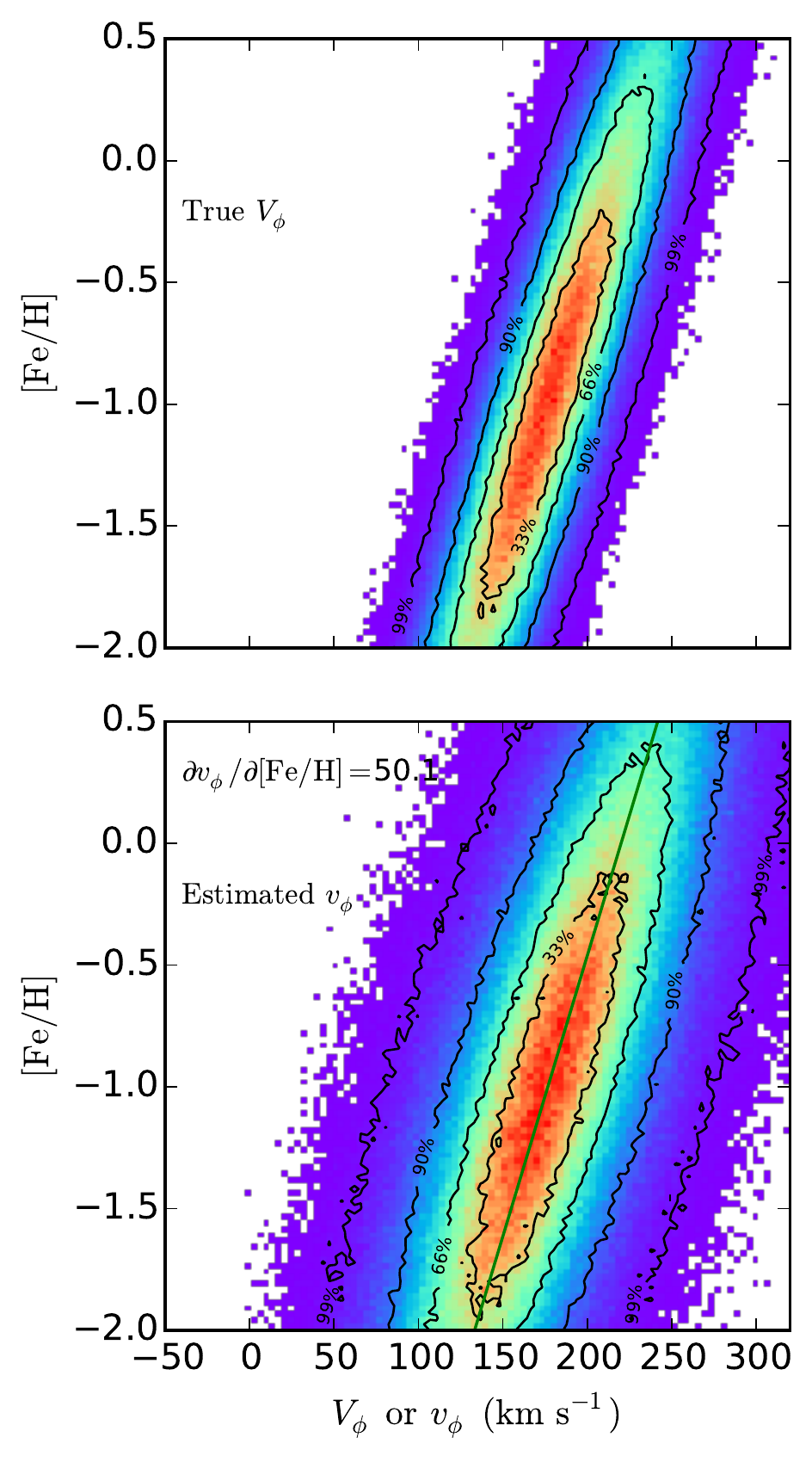}
\caption{ {\bf Top:} Density plots illustrating the relations between true azimuthal velocity and metallicity encoded in our simulated thick disc catalogue, for stars having $|\beta|\geq 0.75$. {\bf Bottom:} Density plots of the derived relations between estimated azimuthal velocity, $\vphib$, and metallicity. }
\label{fig:Besancon_correl}
\end{figure}

We will be investigating correlations between kinematics and
chemistry and it is important to confirm that the use of the estimator
$\vphib$ does not erase or otherwise seriously obscure any underlying
relationships.

The basic Besan\c con model assumes that there is no
age-metallicity-azimuthal velocity relation for the old thin disc
(stars older than $3\Gyr$). In addition, all thick
disc stars are assumed to have the same age ($11\Gyr$), as are all halo
stars ($14\Gyr$). Stars in these two components are also assumed by the model to have no 
correlation between their kinematics and their chemical
composition. However, observations of thick-disc stars are consistent
with a correlation between azimuthal velocity and metallicity
\citep[e.g.][]{Spagna10,Kordopatis11b}. We therefore 
modified the metallicity values of the thick disc stars in the Besan\c con model by an amount that depended  on their azimuthal velocities, to be consistent with
the observational trend from \citet{Kordopatis13c}, namely $\partial
\vphi / \partial \feh = 50 \kms\dex^{-1}$. Then, we introduced a scatter around these metallicity values, drawn randomly from a Gaussian distribution of width $0.3$, to reproduce an intrinsic scatter to the relation. The final adopted input trend can be seen in the top panel of Fig.~\ref{fig:Besancon_correl}.

The results obtained for stars having $|\beta|\geq 0.75$, random errors in distance of 30 per cent, and random $\vlos$ errors of $0.2\kms$ (typical values reached by high-resolution surveys) indicate that we do not introduce spurious correlations between  $\vphi$ and $\feh$ for any of the components. In particular, we  recover for the thick disc  an output slope of  $\partial \vphib / \partial \feh = 50.1 \kms\dex^{-1}$, very close to the input value (bottom panel of Fig.~\ref{fig:Besancon_correl}). 
These tests give us confidence in the use of the estimator to analyse real data. 

That said, we report an increased scatter in the thick-disc metallicity-azimuthal velocity relation when the estimator is used, which indicates that the dispersion of $\vphib$ does not equal that of the true azimuthal velocity distribution.  Indeed, $<V_r^2>$, $<V_\phi^2>$,  $<V_z^2>$ being not equal to zero, they hence cannot be ignored when computing the velocity dispersion of $\vphib$ (see Eq.\,\ref{eq:Vgal_sph}). The dispersion of $\vphib$ is therefore a biased estimator of the true dispersion $\vphi$. Methods to obtain an unbiased estimator of $\sigma_\phi$ were  presented in \citet[][see their Sect.\,5]{Morrison90}, but since they require knowledge of the  intrinsic values of $\sigma_r$ and $\sigma_z$, which we do not have,  we will not use them in this paper.

\subsection{Summary of tests: Pros and cons}
These tests with synthetic data have demonstrated the robustness
of the kinematic estimators. The uncertainty in the derived azimuthal velocity
for a single star when using the estimator $\vphib$ is determined by the combination of the
uncertainty in the line-of-sight velocity (which is generally much
lower than that in the tangential velocity) and by how much $|\beta|$ is different from unity. The reduced uncertainty possible  for $|\beta| \gtrsim 0.7$  
allows for more robust definition of correlations and identification of unusual patterns and outliers. 

However, this reduction in uncertainty is at the 
expense of a reduction in sample size, due to the requirement that the
stars  be located along lines-of-sight towards, or against, Galactic
rotation. Nevertheless, as demonstrated in
Appendix~\ref{app:error_mean_velocity}, similar errors in mean trends
can be obtained with the combination of small sample sizes and low
uncertainties as with the combination of large sample sizes and high
uncertainties (the latter being the case when using all-sky samples
and three-dimensional space motions).

Furthermore, while selecting these lines-of-sight omits stars at small
Galactocentric radii $R$ ($R\leq 6.9\kpc$, see Sect.~\ref{sec:application_to_data}),
the rotation fields allow us to probe the kinematics of stars at larger
$R$  with high accuracy and in an
unbiased manner (up to $R\sim 10\kpc$). This is a clear advantage compared to studies that
select a subsample of high-quality targets with good estimations of
distances and small random errors in proper motions (neglecting
possible systematics on these parameters). For example, in their study
with the APOGEE DR10 catalogue, \citet{Anders14} started with a sample
of $\sim20\,000$ stars and after applying cuts to select stars with the most reliable orbital parameters ended with a `gold' sample of fewer than
$4\,000$ stars, with only a handful outside the extended Solar
neighbourhood.

\smallskip

In addition, the use of a more precise azimuthal velocity leads
to a more precise estimate of the value of the Galactocentric radius
of the guiding centre of the star's orbit ($R_g$), assuming that the
epicyclic approximation provides an appropriate description of the
stellar motion in the plane of the Galaxy.   The guiding centre radius is a good proxy for  the mean
orbital radius and is directly determined by the stellar orbital
angular momentum (about the $z-$axis), $L_z$, being the radius of a
circular orbit of angular momentum equal to $L_z$. We thus have $R_g =
L_z/V_c(R_g)$, with $V_c(R_g)$ being the circular velocity at radius
$R_g$.

Estimation of the stellar orbital angular momentum requires an
estimate of the azimuthal velocity; herein lies the importance of the
increased precision of $\vphib$. The joint distribution of angular
momentum (or, equivalently, $R_g$) and chemical abundances is of
particular interest in constraining dynamical processes that modify
the angular momentum distribution and move stars from their birth
places. Radial migration of stars due to 
torques from transient spiral
structure \citep{Sellwood02} or from the overlap of resonances with multiple non-axisymmetric, long-lived non-axisymmetric patterns \citep[e.g., bar+spirals][]{Minchev10} are of particular current interest; \citet{Sellwood02} showed
that the dominant effect of such an interaction at the corotation
resonance is a change of the star's orbital angular momentum without
an accompanying increase in its orbital eccentricity \citep[dubbed \lq
churning' by][]{Schonrich09a}. In contrast, interactions at the inner or outer Lindblad resonances do lead to an increase in orbital eccentricity and radial epicyclic excursions \citep[hence dubbed \lq blurring' by][]{Schonrich09a}. Angular momentum is lost (gained) by stars at the Inner (Outer) Lindblad resonance   
\citep{Lynden-Bell72} but interactions at the corotation resonance dominate the angular momentum changes and hence are the main source of permanent  changes in guiding centre radii and thus radial migration \citep{Sellwood02}.

Investigation of the chemical properties of stars as a function of their guiding centre radius, $R_g$, rather than their observed, current position $R$ therefore provides a means of removing the effects of `blurring'. Additionally, knowledge of the (current) mean orbital radius facilitates the identification of  stars whose chemical abundances indicate migration from  their birth radius.

A proper determination of the guiding centre radii of the stars would require integration of the stellar orbits within an assumed  Galactic potential, which would require  knowledge of the three velocity components of each target (plus distance). However, under the simplifying assumption of  a flat rotation curve for the Galaxy, $R_g$ can be straightforwardly obtained using the estimated azimuthal velocity as: 

\begin{equation}
R_g=\frac{R \cdot \vphib}{V_c}
\label{eq:Rg}
\end{equation}
where $R$ is the observed Galactocentric radius of the star,
$V_c$ is the circular velocity (assumed to equal
$220\kms$, independent of radius\footnote{We checked how sensitive our  results in the sections below are to the rotation curve by  adopting gradients in the rotation curve  of  $\pm5\kms\kpc^{-1}$ (consistent with the local gradient from the Oort constants)  and found  no significant change in the derived $R_g$ for more than 90 percent of our sample \citep[see][for a discussion on possible values of Oort's constants $A$ and $B$]{BlandHawthorn16}.})  
and $\vphib$ is the estimated stellar
azimuthal velocity derived from the line-of-sight velocity. The scaling with current radial position leads one to expect that the error budget should be dominated by the error in distance, which typically  is larger than that in the estimated azimuthal velocity, for the lines-of-sight considered here.

\smallskip

 We now turn, in the following Sections, to the application of these techniques to real data.

\section{Application to large spectroscopic surveys}
\label{sec:application_to_data}
\subsection{Selection of the APOGEE subsample}
\label{sec:selection_apogee}

We now use the stars from APOGEE data release 12
\citep{Majewski15}; this survey preferentially observed stars along
lines of sight close to the Galactic plane, obtaining high-resolution spectra in the infra-red
wavelength range ($1.51\,\mu$m - $1.69\,\mu$m) that provide a precision on
$\vlos$ of the order of $0.1\kms$. The distances that we adopt in this paper are
computed by \citet{Santiago16}, estimated to have a bias smaller than 2 per cent (based on tests with Hipparcos, CoRoT, cluster and simulated stars). These distances have been obtained using an isochrone-fitting method 
based on the spectroscopic stellar parameters inferred by the APOGEE Stellar Parameters and Chemical Abundances Pipeline \citep[ASPCAP,][]{GarciaPerez16} and the 2MASS photometry. 
Galactocentric
velocities are computed assuming: $R_0=8\kpc$, $\vlsr=220\kms$, $(U_0,
V_0, W_0)=(11.1, 12.2, 7.2)\kms$ \citep{Schonrich10}, together with
the UCAC4 proper motions, when needed.
Throughout this paper, we use the APOGEE DR12 calibrated \meta\, values,  the distribution of which will be referred to as the metallicity distribution function (MDF). We derived the  $\afe$ values following \citet{Hayden15}, \ie\  by adding  the uncalibrated global metallicity, $\rm [M/H]$ to the calibrated $\mathrm{[\alpha/M]}$ and then subtracting the uncalibrated \meta.

\smallskip

\begin{figure}
\centering
\includegraphics[width=0.9\linewidth, angle=0]{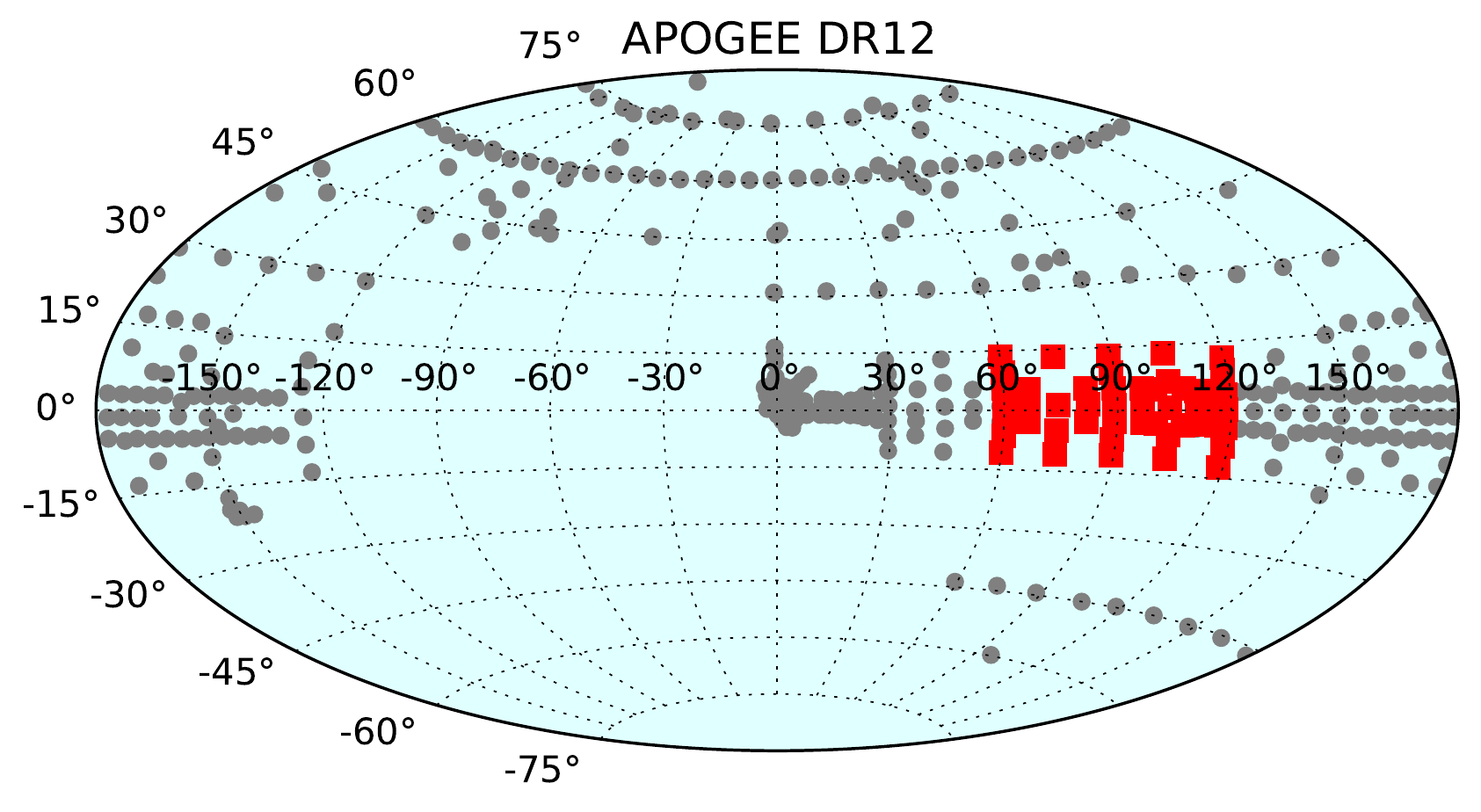}
\caption{Aitoff projection in Galactic coordinates of all the APOGEE DR12 fields that target field stars (filled grey circles). Those fields for which $\vphib$ is a good estimator of $\vphi$ and contain  stars having $\beta \geq 0.8$ are indicated by the red squares. }
\label{fig:APOGEE_aitoff}
\end{figure}

In addition to the quality cuts applied in Table~1 of \citet{Anders14},
we applied these further selection criteria to the catalogue,
following the recommendations of \citet{Holtzman15}: the mean
scatter derived from repeated observations ($V_{\rm scatter}$)
must be $\leq 0.5\kms$ (to remove stars that have a variable radial
velocity); \teff$ \geq 4000\,K$ (since cooler stars
could have under-estimated errors in their derived
parameters)  and errors in line-of-sight distances must be
less than 50 percent (the analysis in Appendix~\ref{sec:rc_validation} demonstrates  the distances used do not have a large enough bias that could have a significant impact on our technique). Furthermore, only field stars are
kept, removing stars belonging to clusters or calibration
fields\footnote{ In practice, this has been achieved by imposing the
length of the string designating the observed field to be equal to six
characters.}.  A total of 61\,207 stars met all the
above criteria. The Aitoff projection in Galactic coordinates
of the fields containing these stars is shown in
Fig.~\ref{fig:APOGEE_aitoff}.  Following the results of Sect.~\ref{sec:Besancon},
stars in lines-of-sight close to the rotation cardinal
direction are selected as follows: $\ell=90^{\circ}\pm30^{\circ}$,
$|b|\leq 25^{\circ}$ (in practice, the maximum $|b|$ is
$\sim15^{\circ}$) and $\beta \geq 0.8$. These fields
are represented as red squares in Fig.~\ref{fig:APOGEE_aitoff}.

\begin{figure}
\centering
\includegraphics[width=0.9\linewidth, angle=0]{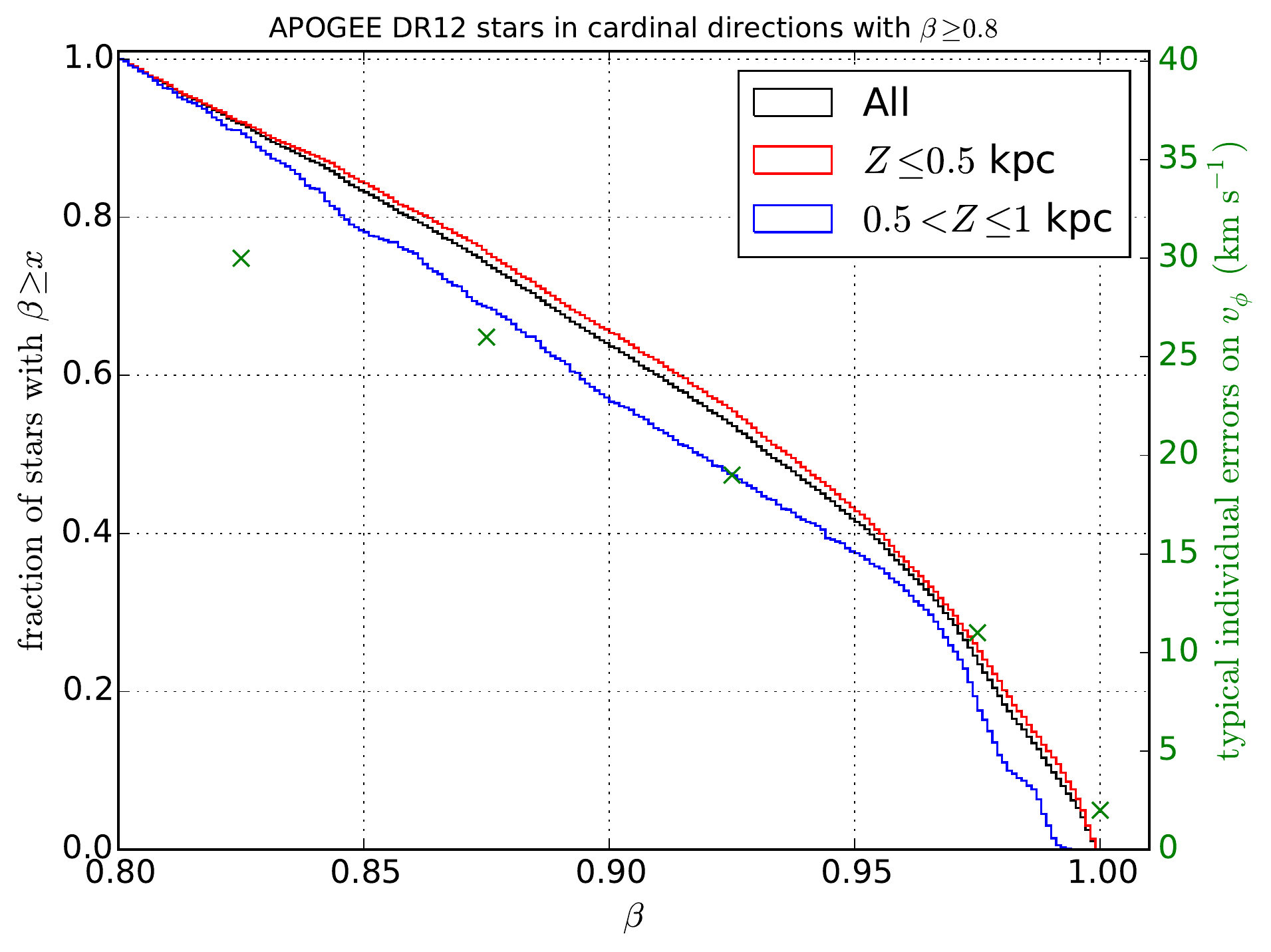} 
\caption{ Cumulative distribution of APOGEE stars towards the cardinal directions, having $\beta$ above a given value (starting from $\beta\geq 0.8$). In black is for the entire sample, in red for the subsample with $|Z|\leq 0.5\kpc$  and in blue for the sub-sample with $0.5<|Z|\leq 1\kpc$. Green `x'-symbols are the associated errors on the individual $\vphib$ of each star, according to Table~\ref{tab:errors_vphi}. We recall, however, that the mean velocities are unbiased (see Sect.~\ref{sec:Besancon}).}
\label{fig:APOGEE_betas_cumul}
\end{figure}

The cumulative histograms of the $\beta$ distribution for the
selected sample is shown in Fig.~\ref{fig:APOGEE_betas_cumul}. From
that plot, one can see that regardless of our cut in the distance from
the plane, $Z$, 60 per cent of our sample has $\beta \gtrsim 0.9$,
corresponding to errors in individual estimated azimuthal velocities
$\Delta{\vphib} \lesssim 20\kms$ (see Table~\ref{tab:errors_vphi_realistic}, columns 2 and 3) and hence uncertainties in guiding centre radii 
$\Delta{R_g} \lesssim1\kpc$. If one now considers 80 per cent of our
sample, then $\Delta{\vphib} \lesssim 30\kms$ and $\Delta{R_g}
\lesssim1.8\kpc$.

\begin{figure}
\centering
\includegraphics[width=0.9\linewidth, angle=0]{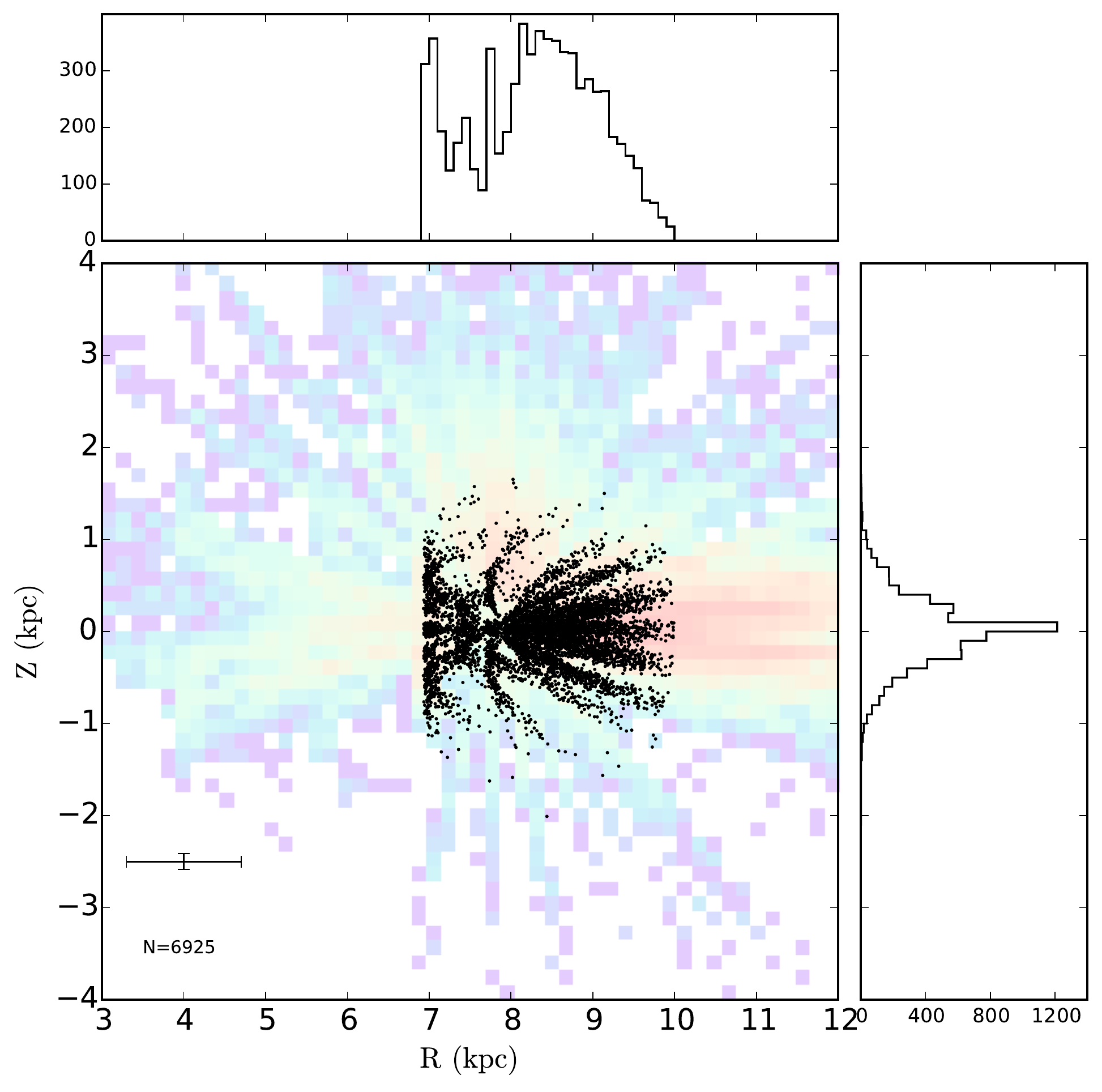} 
\caption{Full APOGEE sample (underlying 2D histogram) and sub-sample towards the cardinal directions (black circles), for the stars fulfilling our selection criteria. The Sun is located at $(R,Z)=(8,0)\kpc$. Typical error bars on the positions for the entire sample are plotted in the bottom left corner. Side histograms show the distribution in $R$ and $Z$ for the sub-sample towards the cardinal directions.} 
\label{fig:Galactic_positions}
\end{figure}

\begin{figure}
\centering
\includegraphics[width=0.9\linewidth, angle=0]{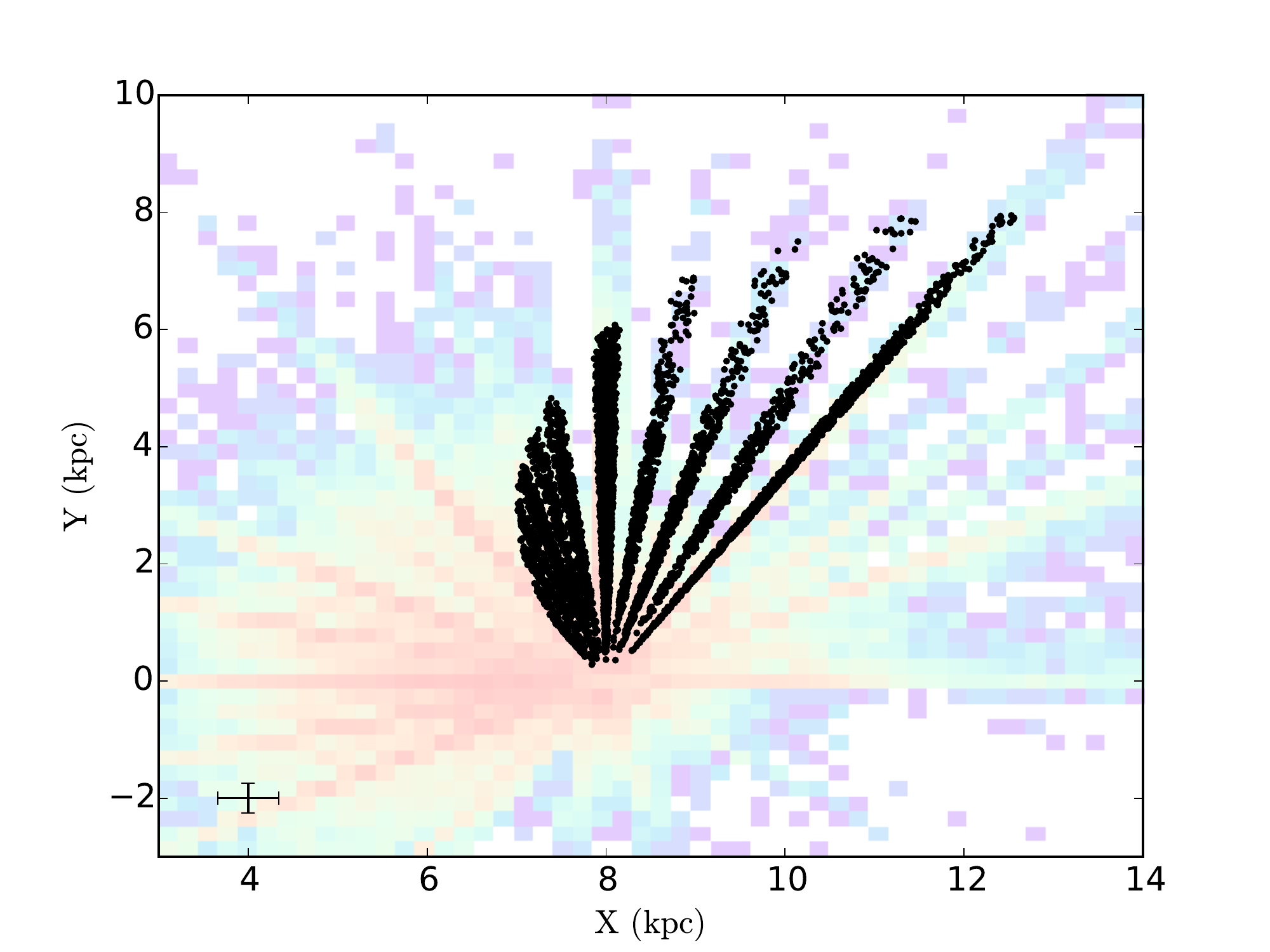} 
\caption{  As Fig.~\ref{fig:Galactic_positions}, but showing the stars in a Galactocentric Cartesian coordinate system $XY$. The Sun is located at $(X,Y)=(8,0)\kpc$ and the direction of Galactic rotation is towards  positive $Y$.}
\label{fig:Galactic_positions_XY}
\end{figure}

 The spatial distribution of the sample selected using our
criteria is shown in Fig.~\ref{fig:Galactic_positions} using
cylindrical coordinates $(R,Z)$ and in
Fig.~\ref{fig:Galactic_positions_XY} using cartesian coordinates
$(X,Y,Z)$. One can see from Fig.~\ref{fig:Galactic_positions} that
the criterion in longitude has restricted the inclusion of stars at small 
Galactocentric radii (i.e. in the inner Galaxy). Indeed, 
the smallest value for $R$ for this  sample is $6.9\kpc$ (as expected from Eq.~\ref{eq:u2}, adopting the median  heliocentric distance of $3\kpc$ and our lowest longitude, $60^\circ$).
The selected sub-sample spans therefore $6.9<R<10\kpc$ and $|Z|<1.5\kpc$, as seen from the side histograms of Fig.~\ref{fig:Galactic_positions}.

\begin{figure}
\centering
\includegraphics[width=0.9\linewidth, angle=0]{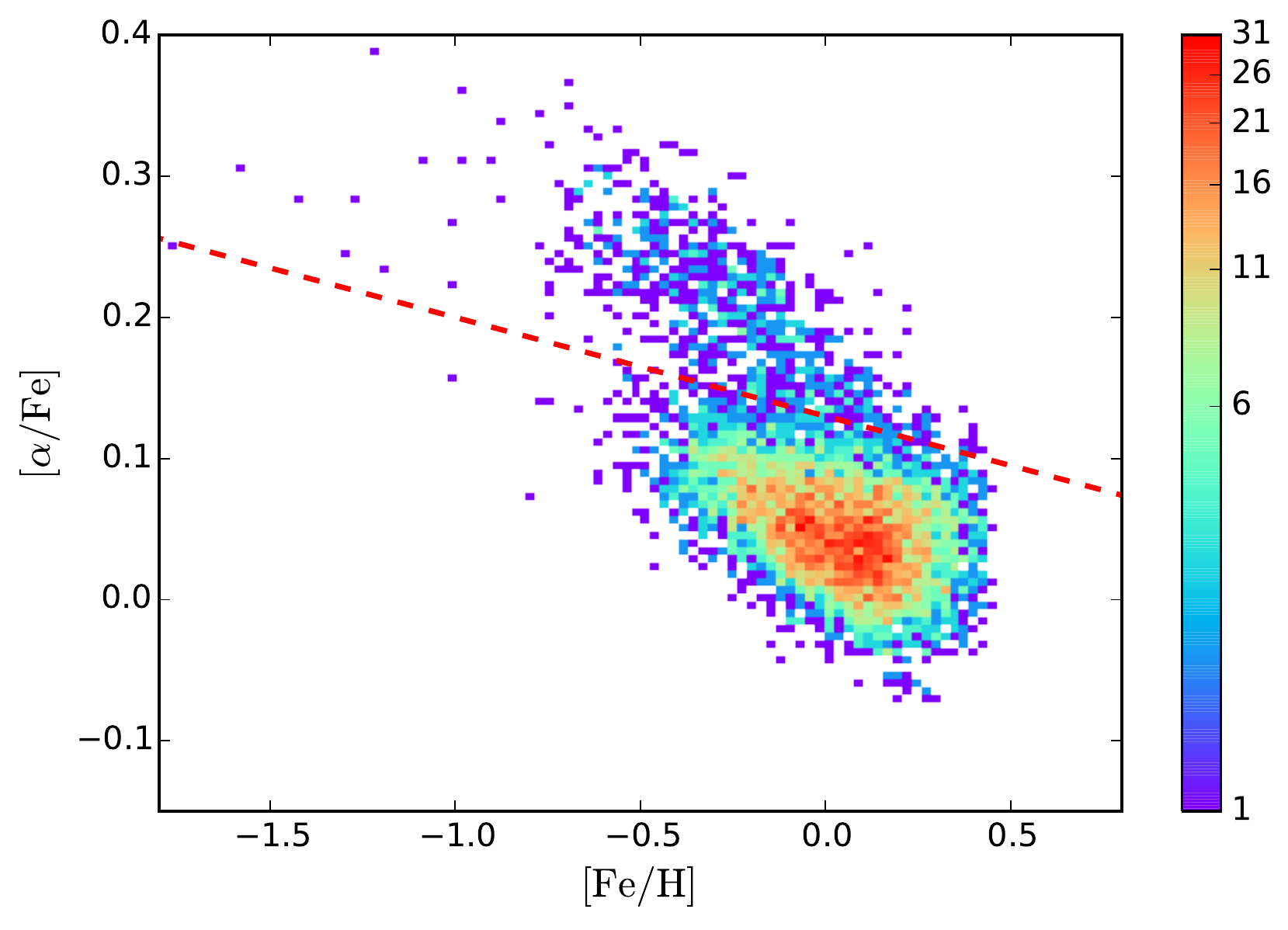}
\caption{ Density plot showing the $\alpha$ abundances relative to iron as a function of iron abundance for the stars towards the cardinal directions. The red line can be written as $\afe=-0.07\cdot \feh+0.13$, and has been chosen in order to separate the $\alpha$-high and the $\alpha$-low populations throughout the paper.  }
\label{fig:alphafe_cardinal}
\end{figure}

 Finally, Fig.~\ref{fig:alphafe_cardinal} shows the $\alpha$
 abundances relative to iron as a function of \feh\, for the stars
 that have passed our selection criteria. The slope and normalisation
 of the line have been chosen by eye as a convenient separation of the
 $\alpha$-high and the $\alpha$-low populations. This
 simple cut divides the stars into two groups over most of the
 metallicity range.  However, we emphasise that the separation into two sequences is more ambiguous at the metal-rich end where the two populations may merge \citep[see, for
 example,][]{Nidever14}. \citet{Kordopatis15b} separated the sequences statistically, 
 using all-sky data from the Gaia-ESO survey, and the simple cut here captures the important aspects.

\subsection{Estimation of the azimuthal velocities in APOGEE}

\begin{figure}
\centering
\includegraphics[width=0.9\linewidth, angle=0]{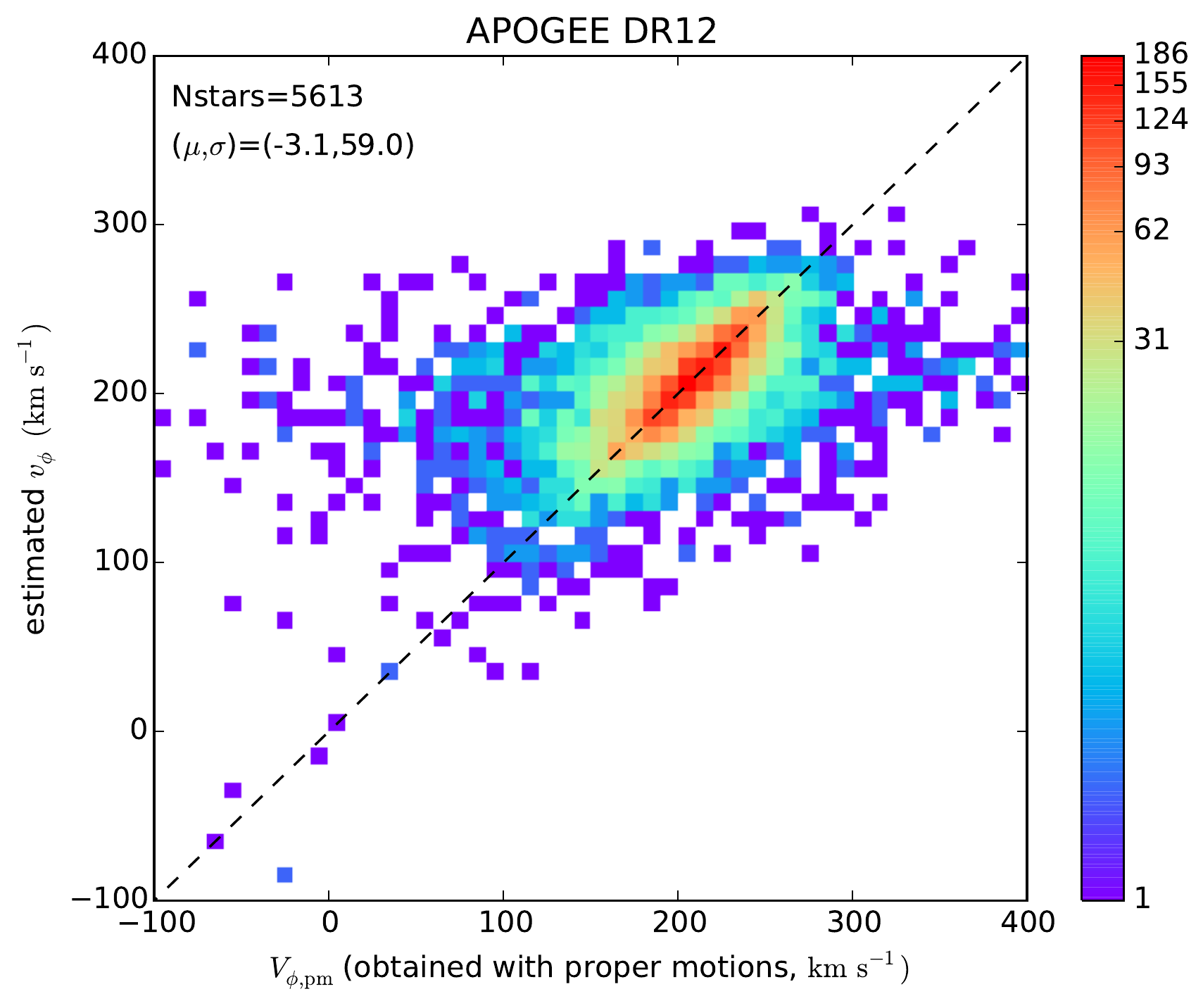}
\caption{Density plot showing the comparison between the $\vphib$ estimator and the derived $\vphipm$ using UCAC4 proper motions for the APOGEE targets. Targets are selected with $|b| < 25^{\circ}; \beta >0.8$, S/N$>60$ and $\ell \pm 30^{\circ}$ from the rotation direction.}
\label{fig:APOGEE_all_vphi}
\end{figure}

In this section we assess the quality of the azimuthal velocities using the APOGEE data. Figure~\ref{fig:APOGEE_all_vphi} shows the comparison of $\vphipm$ (the azimuthal velocity computed using the proper motions) and $\vphib$ (the azimuthal velocity estimated without the proper motions), for the stars towards the direction of Galactic rotation, as selected. 
One can see that the velocities overall agree relatively well.  The bias is $-3.1\kms$ and the dispersion is $59\kms$. As will be shown below,  this relatively large value for the dispersion is primarily due to the high uncertainties in the proper motions, which enter when computing $\vphipm$. 

\begin{figure}
\centering
\includegraphics[width=0.9\linewidth, angle=0]{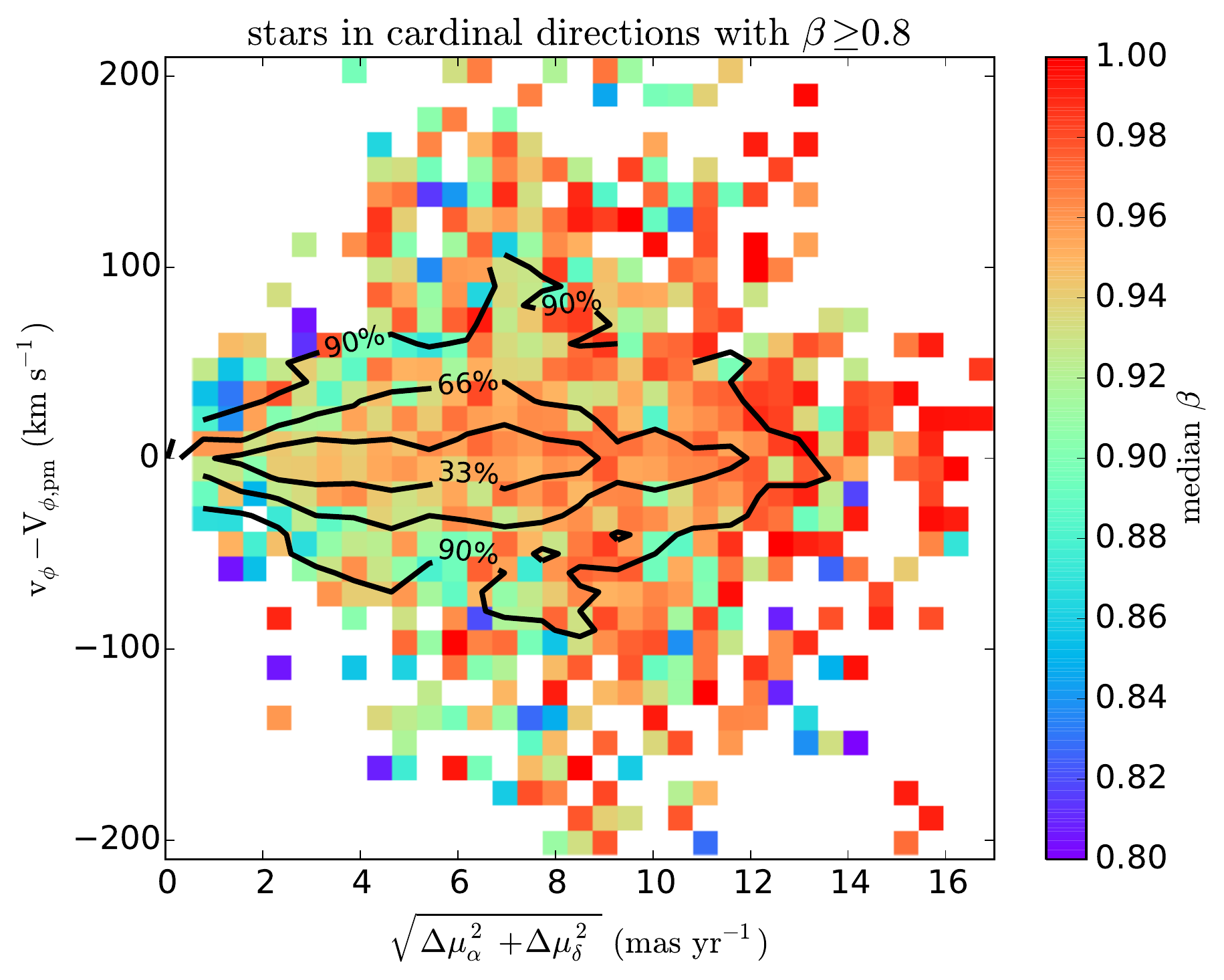}
\caption {The difference between the $\vphib$ estimator and the
derived $\vphipm$ using UCAC4 proper motions, as a function of the
published errors in proper motions, evaluated as a quadratic sum of
the errors in right ascension ($\Delta \mu_\alpha$) and declination
($\Delta \mu_\delta$). The $y$-axis has been truncated at $\pm210\kms$
for reasons of clarity. The colour-coding corresponds to the
median $\beta$ per bin in this parameter space. The
contours marked $33, 66$ and $90$\,per\,cent indicate the
overall distribution of the sample in this plane. The largest discrepancies between the two velocity
estimators are not necessarily found for those stars with the largest
proper motion errors or with the lowest $\beta$. This implies
that it is not possible to select a sample with high-quality 3D space
motions simply by pruning on $\Delta \mu_\alpha$ or
$\Delta \mu_\delta$.}

\label{fig:APOGEE_all_vphi_pmerrors}
\end{figure}

\begin{figure}
\centering
\includegraphics[width=0.9\linewidth, angle=0]{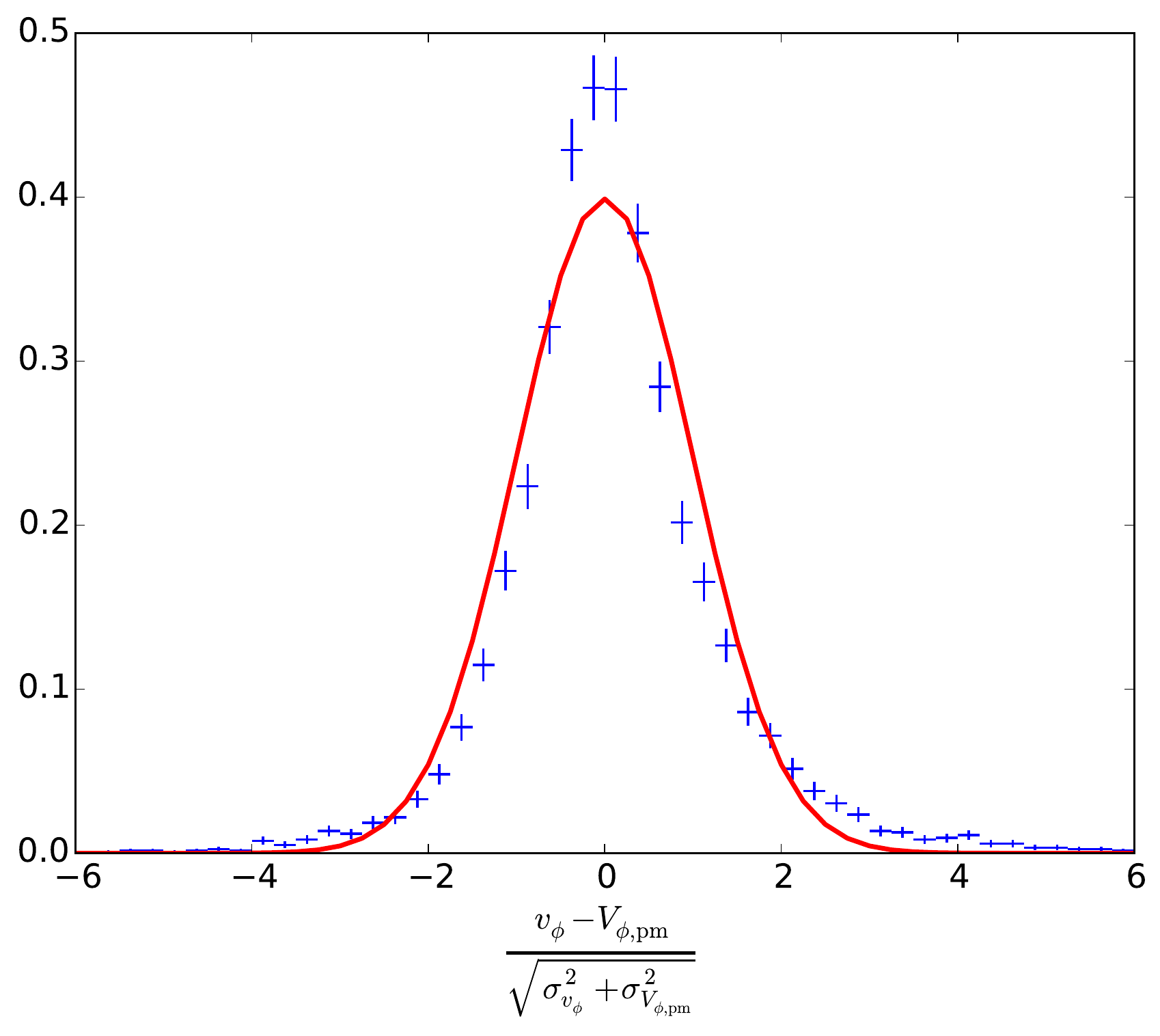}
\caption {Histogram of the difference between $\vphib$ and $\vphipm$, normalised by the quadratic sum of the errors of both velocity estimations. The horizontal error bars mark the widths of the bins and the vertical error bars indicate Poisson
uncertainties. The full curve is a Gaussian of zero mean and unit dispersion, not a fit to the data.}
\label{fig:APOGEE_statistical_test_errors}
\end{figure}

\begin{figure*}
\centering
\includegraphics[width=0.85\linewidth, angle=0]{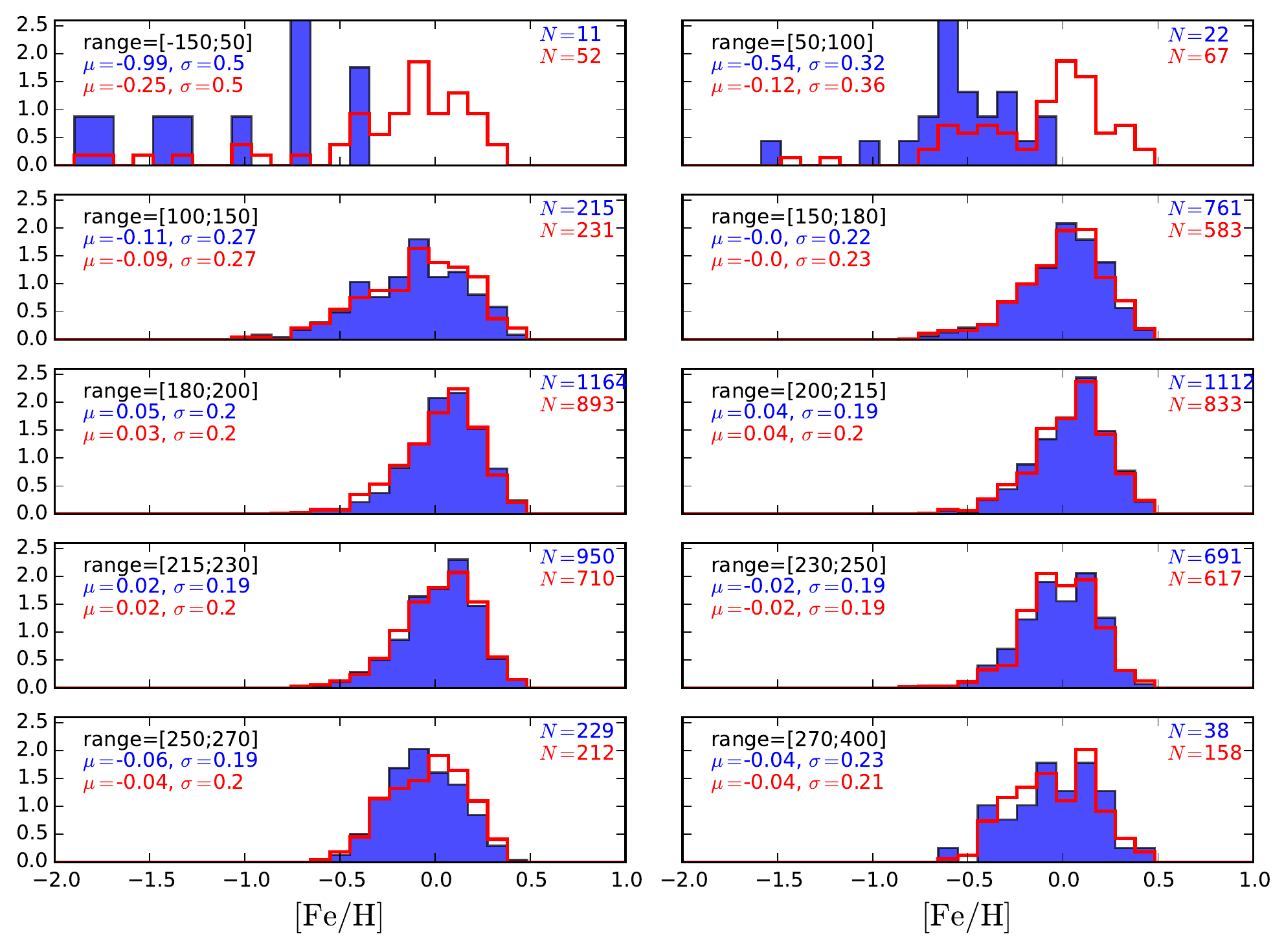}
\caption{Normalised metallicity histograms for stars within different azimuthal velocity ranges (as given  in the upper lefthand corner of each box) and $7\leq R \leq 9\kpc$ ; $|Z|\leq 1\kpc$. The filled blue histograms show the distributions for  the $\vphib$ estimator while the red histograms correspond to the  $\vphipm$ estimator.  The mean and dispersion of each distribution are also indicated within each panel. For the lowest azimuthal velocities (top panels) the red and blue histograms are in stark disagreement. In particular, the high metallicity values reached for the red histograms (obtained with $\vphipm$) indicate that the velocities obtained with proper motions are subject to large errors, since it is not expected that super-metal rich stars would have such low values of orbital angular momentum or indeed be counter-rotating.}
\label{fig:APOGEE_MH_vphi_bins}
\end{figure*}

Evaluation of the differences between the two velocity estimations as a function of the published UCAC4 proper motion errors (denoted $\Delta \mu$) and the $\beta$ values (Fig.~\ref{fig:APOGEE_all_vphi_pmerrors}) shows that the stars with the lowest $\beta$ values in our sample ($\beta=0.8$) are only on a few occasions the ones with the largest differences in the two velocities. Indeed, while this might be the case  for the stars with low $\Delta\mu$ (for example, the blue pixels at around $3 \mas\yr^{-1}$ and $\pm 80 \kms$), this  is certainly not the case for those stars with     $\Delta\mu\gtrsim 5\mas\yr^{-1}$ (corresponding to transverse velocity errors of $\sim 75\kms$ at $3\kpc$), where stars with $\beta\gtrsim0.95$ and $|\vphib-\vphipm| \gtrsim150\kms$ are found.
Furthermore, the underlying distribution of the stars in Fig.~\ref{fig:APOGEE_all_vphi_pmerrors}, seen from the contour lines, indicates that the targets with $|\vphib-\vphipm| \gtrsim150\kms$ are too frequent to be  $2-3\sigma$ outliers of the error distribution in proper motions.

In order to investigate whether the large scatter seen in Fig.~\ref{fig:APOGEE_all_vphi_pmerrors} can be due to the combination of errors in distance, $\vlos$, proper motion and $\vphib$, we compute in Fig.~\ref{fig:APOGEE_statistical_test_errors}  the discrepancies between the two velocity estimators,  normalised by the quadrature sum of the errors of $\vphipm$ and $\vphib$ (in blue). In the case where the central limit theorem applied to the
data, each velocity estimator were unbiased and that each error is properly estimated\footnote{We note, however, that the errors in distance and hence in velocity are not Gaussian, like assumed here.}, the histogram would hence be a Gaussian of unit dispersion. This
expectation is roughly met, but  the tails of the distribution are wider than in the Gaussian case. Given that the tests done in Sect.~\ref{sec:Besancon} indicate that distance biases, or other assumptions in the computation of $\vphib$, are not introducing significant offsets to our velocity estimator, these tails are attributed to underestimated proper motion errors (and/or under-estimated distance errors).

\smallskip 

We further investigated the different velocity estimators where the results are the most discrepant: 
at low $\vphipm$ values,  where we find $\vphib\sim200\kms$ for $\vphipm\sim0\kms$, and at high $\vphipm$ values, where we find $\vphib\lesssim250\kms$ for $\vphipm\gtrsim350\kms$.
In particular, to assess the nature of these stars, we investigated the metallicity distribution functions (MDF) for different ranges  in azimuthal velocity, defined by  either the $\vphipm$ values or the $\vphib$ values. The volume for this comparison is restricted to the extended Solar neighbourhood ($7\leq R \leq 9\kpc$ and $|Z|\leq 1\kpc$).

The iron abundance  histograms shown in Fig.~\ref{fig:APOGEE_MH_vphi_bins} indicate that there is a fair agreement of the MDFs for stars selected to have velocities (either $\vphipm$ or $\vphib$) between $100\kms$ and $250\kms$. At lower velocities (i.e.~stars on lower angular momentum orbits), the MDF obtained from  a selection on $\vphipm$ exhibits the presence of a metal-rich (solar and super-solar values) and $\alpha$-low population, 
which is highly unlikely to be real, given the velocity dispersions and MDF of the thin disc, thick disc and halo \citep[see for example,][ and references therein]{Chiba00, Soubiran03,Carollo10}. 

\begin{figure*}
\centering
\includegraphics[width=0.85\linewidth, angle=0]{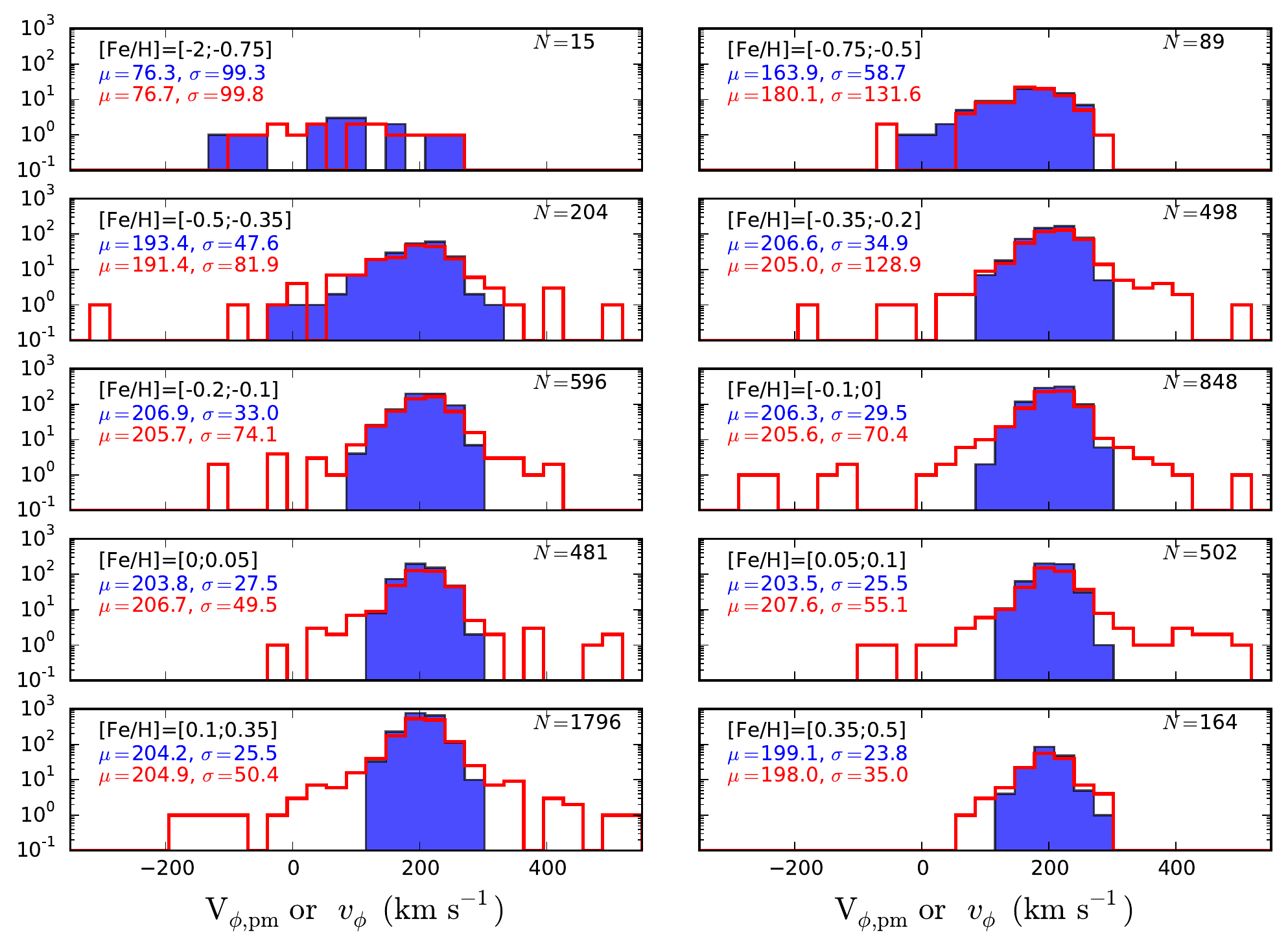}
\caption{ Azimuthal velocity histograms (in logarithmic scale) for different \meta\, ranges (the exact range is written in the upper left corner of each box). The filled blue histograms show the distributions for $\vphib$, the red histograms correspond to $\vphipm$.  The means and dispersions of each distribution are also indicated within each panel. } 
\label{fig:APOGEE_MH_vphi_bins2}
\end{figure*}

 Figure~\ref{fig:APOGEE_MH_vphi_bins2} compares histograms of the distributions  of  $\vphib$ and $\vphipm$, for subsets of the sample selected  in different $\feh$ bins (note the logarithmic scale on the $y$-axes). It is apparent  that in each $\feh$ bin, with the exception of the most metal-poor one, the distributions of $\vphipm$ are much broader that those of $\vphib$, extending from negative values - unexpected for disc stars - up to higher than $+450\kms$. 
 In light of the results of Fig.~\ref{fig:APOGEE_statistical_test_errors} and \ref{fig:APOGEE_MH_vphi_bins}, we  interpret this broadening as reflecting the  uncertainties and systematics  in the proper motions. The numerical dominance of the higher metallicity  stars  in the observed sample means that they will also dominate those with erroneously high velocities derived from the proper motions.

\begin{figure*}
\centering
\includegraphics[width=0.85\linewidth, angle=0]{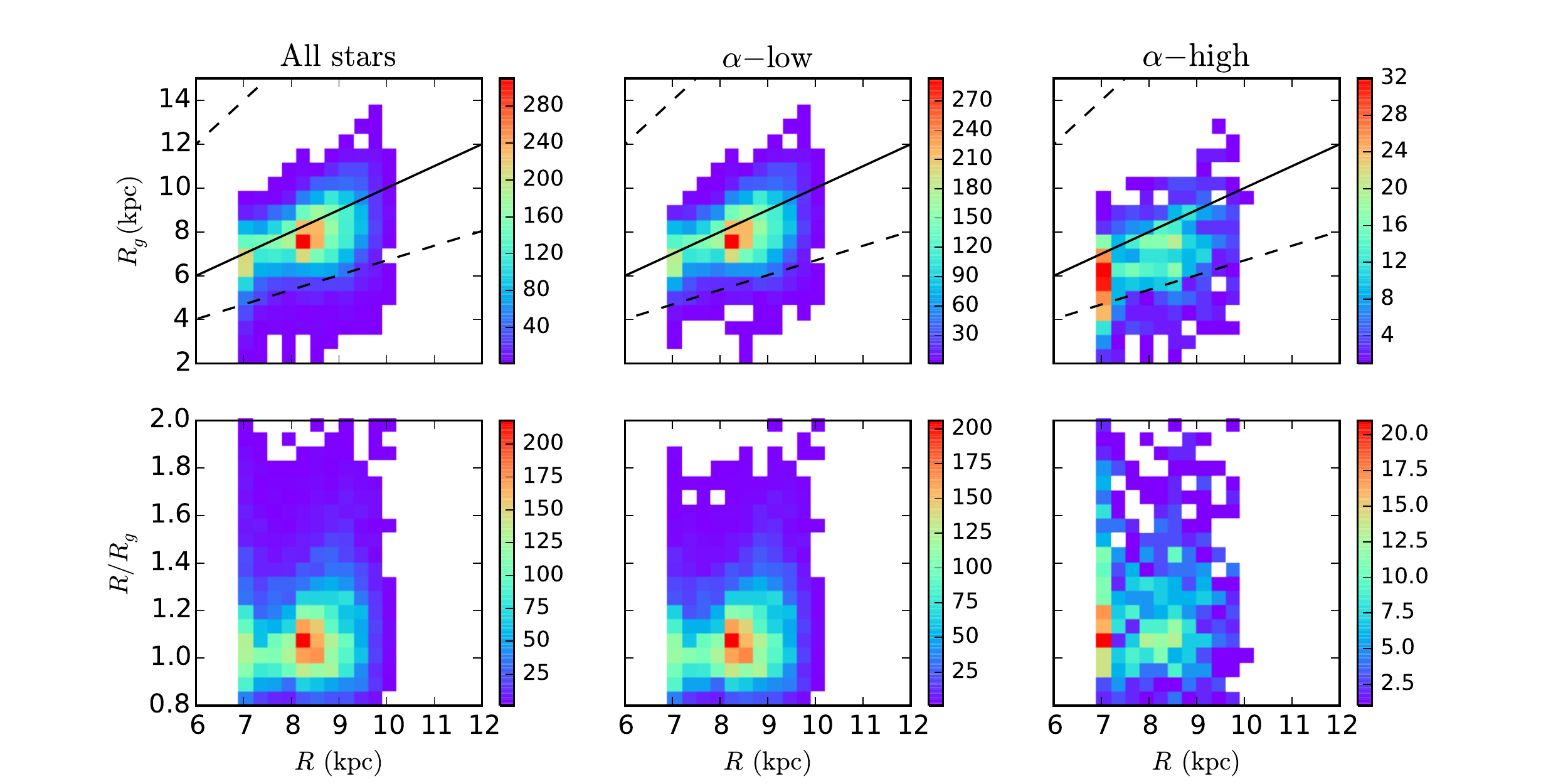}
\caption{2D histograms of $R_g$ vs $R$ (top row) and $R_g/R$ vs $R$ (bottom row) for all the APOGEE stars we selected towards Galactic rotation (left column), the $\alpha$-low sequence (middle column), and the $\alpha$-high sequence (right column). The black solid line represents circular orbits. The dashed lines represent the combination of $R_g$ and $R$ for which $e\geq0.5$. The drop in number density, indicated by the colour-coding, at around $R\sim 8 \kpc$ is related purely to the target selection function.   }
\label{fig:APOGEE_Rg_vs_R}
\end{figure*}

Finally, the plots of the left column of Fig.~\ref{fig:APOGEE_Rg_vs_R}
show the estimated guiding centre radii (top) and and the ratio
$R/R_g$ (bottom) as a function of the observed Galactocentric radial
coordinates of the stars. One can see that most of the stars have $R_g
\lesssim R$.  This is related to the fact that in an axisymmetric
system, the effective potential rises more steeply moving inwards from $R_g$ than it does moving outwards,
resulting in stars, as they oscillate about $R_g$,
spending more time beyond $R_g$ than they do interior to it \citep[see
for example,][]{Schonrich12b}. The exponential nature of the disc
stellar density profile also means that at any radius there are more
stars interior to that radius (to be observed in an excursion beyond
their guiding centre radius) than exterior to that radius (to be
observed in an excursion interior to their guiding centre radius),
strengthening the expectation to observe more stars with $R_g < R$.
The black dashed lines in Fig.~\ref{fig:APOGEE_Rg_vs_R} show the
boundaries beyond which stars have an orbital
eccentricity\footnote{Since no values for the apocentre and pericentre
are known for the stars of our sample, we estimated the lower limit of
the eccentricity as: $e > |R- R_g|/R_g$.} $e\geq 0.5$ and thus the
epicycle approximation may be a poor assumption of the orbit
\citep[however, see][for a discussion of different levels of
approximation]{Dehnen99}. We checked that our results do not change
whether or not we retain those stars beyond the boundaries (primarily
thick-disc stars with $R_g\lesssim 4-5\kpc$).

To summarise, the results of this section indicate that:
(i) for some stars the quoted proper motion (and/or distance) uncertainties are
under-estimated, (ii) there is no simple way to select a clean
sample based on low quoted $\Delta\mu$ and $\Delta d$ in order to obtain high-quality 3D
velocities and (iii) one should pay particular attention when
trying to identify kinematically peculiar stars based on velocities
derived using the proper motions.

\section{$\aM$ vs $R_g$ and $\vphib$ plots for the stars in the cardinal directions.}
\label{sect:chemodynamics}

\begin{figure*}
\begin{center}
\includegraphics[width=0.85\textwidth]{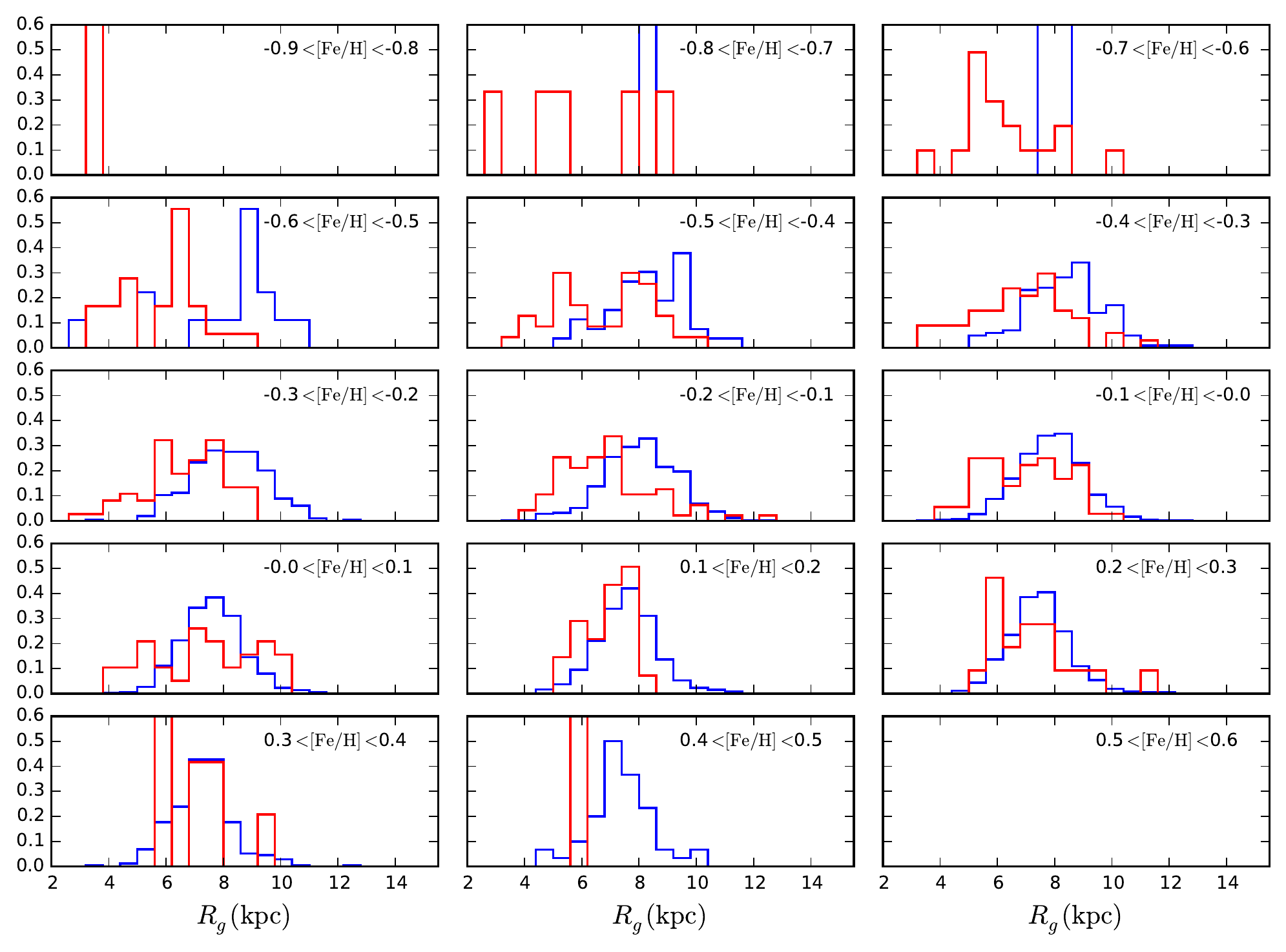}
\caption{Normalised histograms of the $R_g$ distribution of the $\alpha$-high (in red) and $\alpha$-low (in blue) populations closer than $0.5\kpc$ from the plane (with no constraint in radial coordinate), in $0.1\dex$ wide bins of iron abundance. The separation is done according to the line defined in Fig.~\ref{fig:alphafe_cardinal}. Note that the bottom right plot is blank, as there are no stars with this high an iron abundance.}
\label{fig:Rg_meta_bins}
\end{center}
\end{figure*}

In this section we investigate the correlation between $\vphib$ (and/or the derived guiding radius) and the chemical composition of the stars.
Figure~\ref{fig:Rg_meta_bins} shows the range of guiding radii for stars in narrow  bins of $\meta$. 
 We recall that uncertainties in $R_g$ are mainly dominated by the errors in the estimated radial position (see Eq.\,\ref{eq:Rg}). We restrict our selection of stars to include only those within $0.5\kpc$ of  the Galactic plane. This leads to  median errors in $R_g$ of the order of $1\kpc$, based on the distance error estimates of \citet{Santiago16}. The red and blue histograms in Fig.~\ref{fig:Rg_meta_bins} represent the results for the $\alpha-$high and $\alpha-$low populations, respectively, where these two populations  are defined relative to  the red line plotted in Fig.~\ref{fig:alphafe_cardinal}.  The present sample probes a significantly larger volume than  the high-resolution studies of bright stars in the solar neighbourhood for which very precise kinematics are available from the Hipparcos satellite, and a more numerous sample, by around a factor of ten.

One can see that the two populations exhibit very  different distributions in $R_g$ for iron abundances up to the  Solar value, albeit that both show unimodal distributions. In particular, at any given $\feh$ the $\alpha$-high stars have smaller guiding centre radii  than do the $\alpha$-low population. Recall, however, that our sample is biased towards the outer disc, so that the results of Fig.~\ref{fig:Rg_meta_bins} do not imply that there are no $\alpha$-low stars with small guiding radii. On the other hand, Fig.~\ref{fig:Rg_meta_bins} does imply that $\alpha$-high stars are less numerous in the outer Galaxy \citep[as already suggested by previous studies, see for example][]{Bensby11, Cheng12b, Bovy12b, Haywood13}. 
A more general conclusion is that  the dynamical mechanisms responsible for the formation (and hence the spatial distribution) of the thin and thick discs have differed throughout the entire history of the Milky Way (since the stellar kinematics and their correlation with chemistry reflect the evolutionary processes experienced by a population).
Furthermore, while the lowest metallicity $\alpha$-low stars (below $\feh\sim~-0.4$) have a wide range of guiding centre radii, the peak of the distribution is  around $R_g\sim9\kpc$. This implies that the vast majority of these stars are not just visiting the extended Solar neighbourhood through epicyclic excursions,  but were either born within $1\kpc$ from the Sun, or have radially migrated from more distant regions of the disc.

\begin{figure*}
\begin{center}
\includegraphics[width=\textwidth]{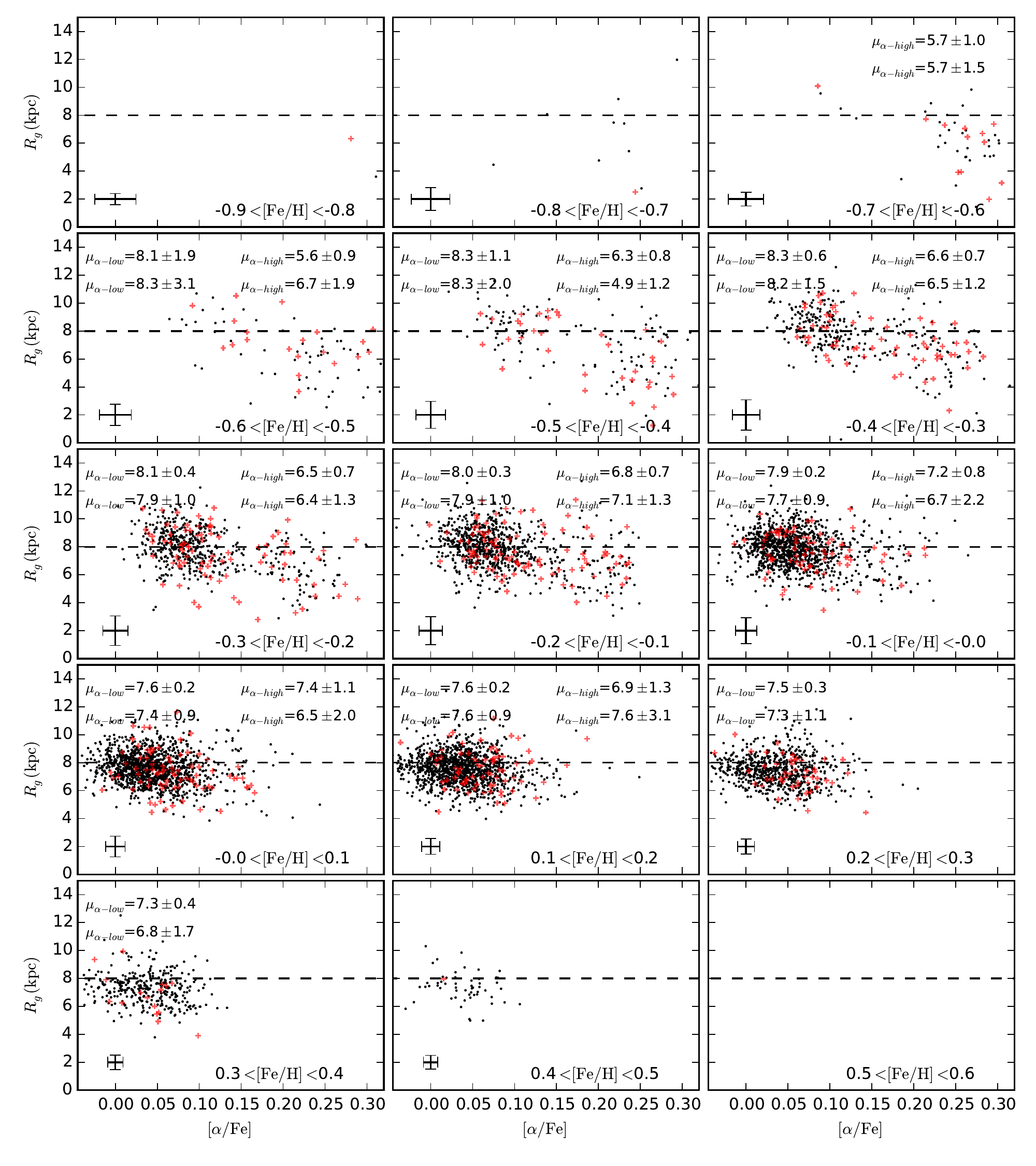}
\caption{Filled black circles and '$+$' red symbols correspond to stars within $[0-0.5]\kpc$ and $[0.5-1]\kpc$ from the Galactic plane. Mean error bars are plotted in the lower-left corner of each box. 
The top-left and top-right corners of each box contain  the mean $R_g$  for each of the  $\alpha$-low and $\alpha$-high populations, respectively. The values on the first line represent the mean $R_g$ for the samples close to the plane, and the values in the second line represent the mean $R_g$ for the samples farther from the plane, provided each sample has at least five stars.}
\label{fig:Rg_meta_bins_scatter}
\end{center}
\end{figure*}

Figure~\ref{fig:Rg_meta_bins_scatter} shows  the scatter plots corresponding to the histograms of Fig.~\ref{fig:Rg_meta_bins}. 
The two populations are clearly separated in these plots (manifest as the two groups of stars, in particular at low iron-abundances), with again the   $\alpha$-high stars more dominant in the inner Galaxy.  We further divided each $\alpha$-sequence into  stars closer and further that 500~pc from the Galactic mid-plane (those stars at distances more than $0.5\kpc$ from the plane are identified by red `$+$' symbols).

\begin{table*}
\caption{$p$-values at different metallicity bins of the Kolmogorov-Smirnov tests for the $R_g$ distribution of the $\alpha$-low and $\alpha$-high stars closer and farther than $0.5\kpc$ from the plane}
\begin{center}
\begin{tabular}{|c|ccccccccccc|}
\hline \hline
Centre of the $\feh$ bin & $-0.65$ & $-0.55$ & $-0.45$ & $-0.35$ & $-0.25$ & $-0.15$ & $-0.05$ & $0.05$ & $0.15$ & $0.25$  & $0.35$   \\ \hline      
$p$-value $\alpha$-low 	& -		& 0.7	& 0.63	& 0.68	& 0.45	& 0.05 	& 0.73 	& 0.1	& 0.14	& 0.16	& 0.29\\
$p$-value $\alpha$-high 	& 0.96 	& 0.3 	& 0.01 	& 0.31	&0.44	& 0.66	& 0.3 	& 0.0	& 0.07	& -		& - \\ \hline
\end{tabular}
\end{center}
\label{tab:KS_tests}
\end{table*}%

 The mean $R_g$ for each $\alpha$-sequence close to, and far from, the plane are reported in each panel of Fig.~\ref{fig:Rg_meta_bins_scatter}. Furthermore, 
the $p$-values derived from the two-sample Kolmogorov-Smirnov tests between the $R_g$ distribution of the sub-samples closer and farther from the plane, for each $\alpha$-sequence, are reported in Table~\ref{tab:KS_tests}. On this basis ($p$-values$\gtrsim 0.3$ for most of the metallicity bins and similar mean $R_g$ for each population close and far from the plane), we cannot reject the null hypothesis  that stars closer and farther than 500~pc from the plane   have the same $R_g$ distributions. 
Despite being difficult to draw firm conclusions on the properties of the vertical velocity dispersions of the populations due to degeneracies with their radial profiles, our results are consistent with both $\alpha$-high and $\alpha$-low populations  being stratified vertically and having quasi- isothermal/constant vertical  velocity dispersions \citep[see also][]{Binney12b, Binney14b}.

\begin{figure*}
\begin{center}
\includegraphics[width=0.9\textwidth]{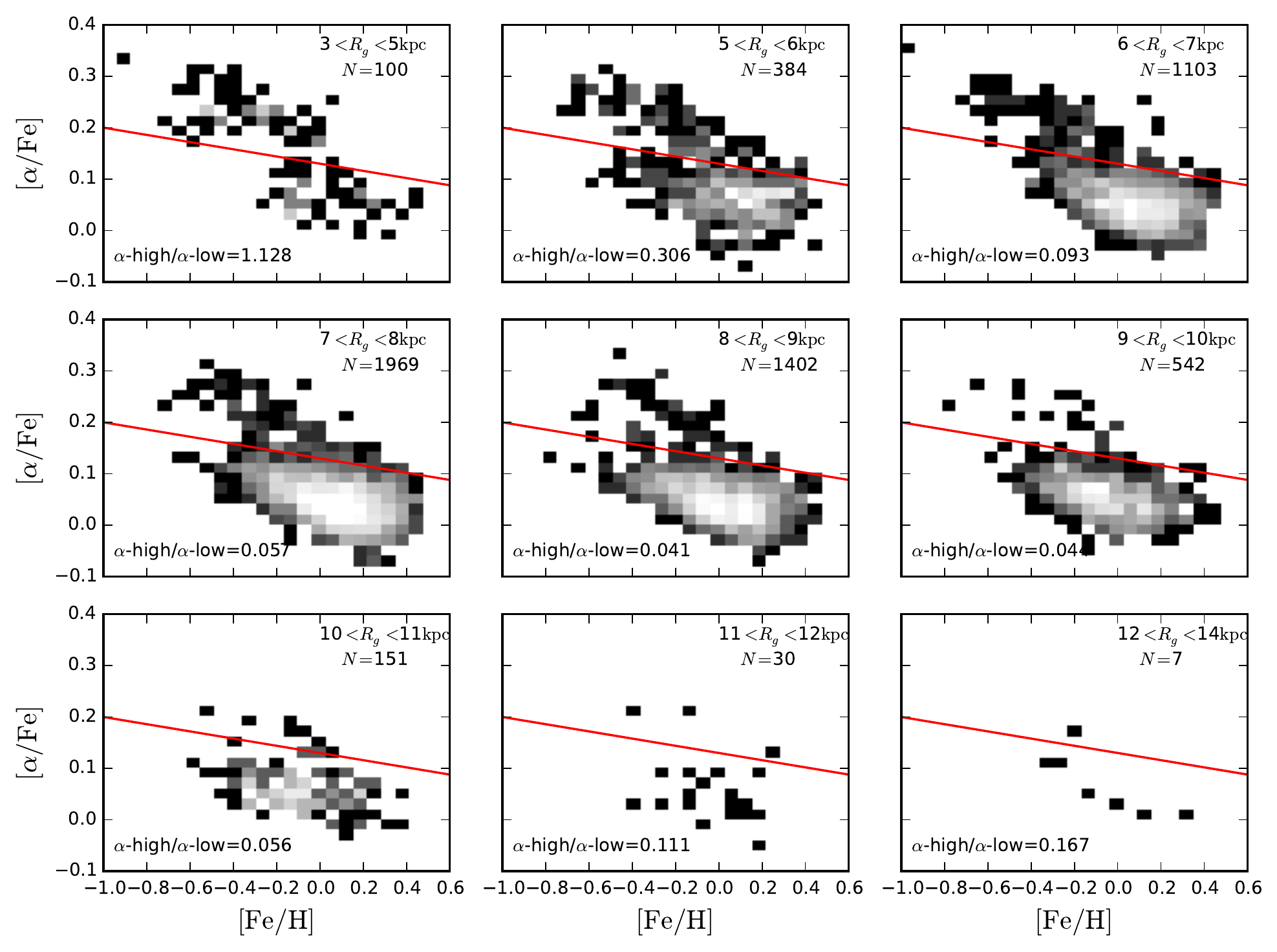}
\caption{$\aM$ as a function of \meta\, for different ranges of guiding radii, for stars with $|Z|\leq 0.5\kpc$. The red dividing line from Fig.~\ref{fig:alphafe_cardinal} is reproduced on each panel. The change in the ratio of the $\alpha$-high to the $\alpha$-low populations as a function of $R_g$ is plotted in Fig.~\ref{fig:Population_ratio}.}
\label{fig:alpha_meta_Rgbins}
\end{center}
\end{figure*}

 \begin{figure}
\begin{center}
\includegraphics[width=0.45\textwidth]{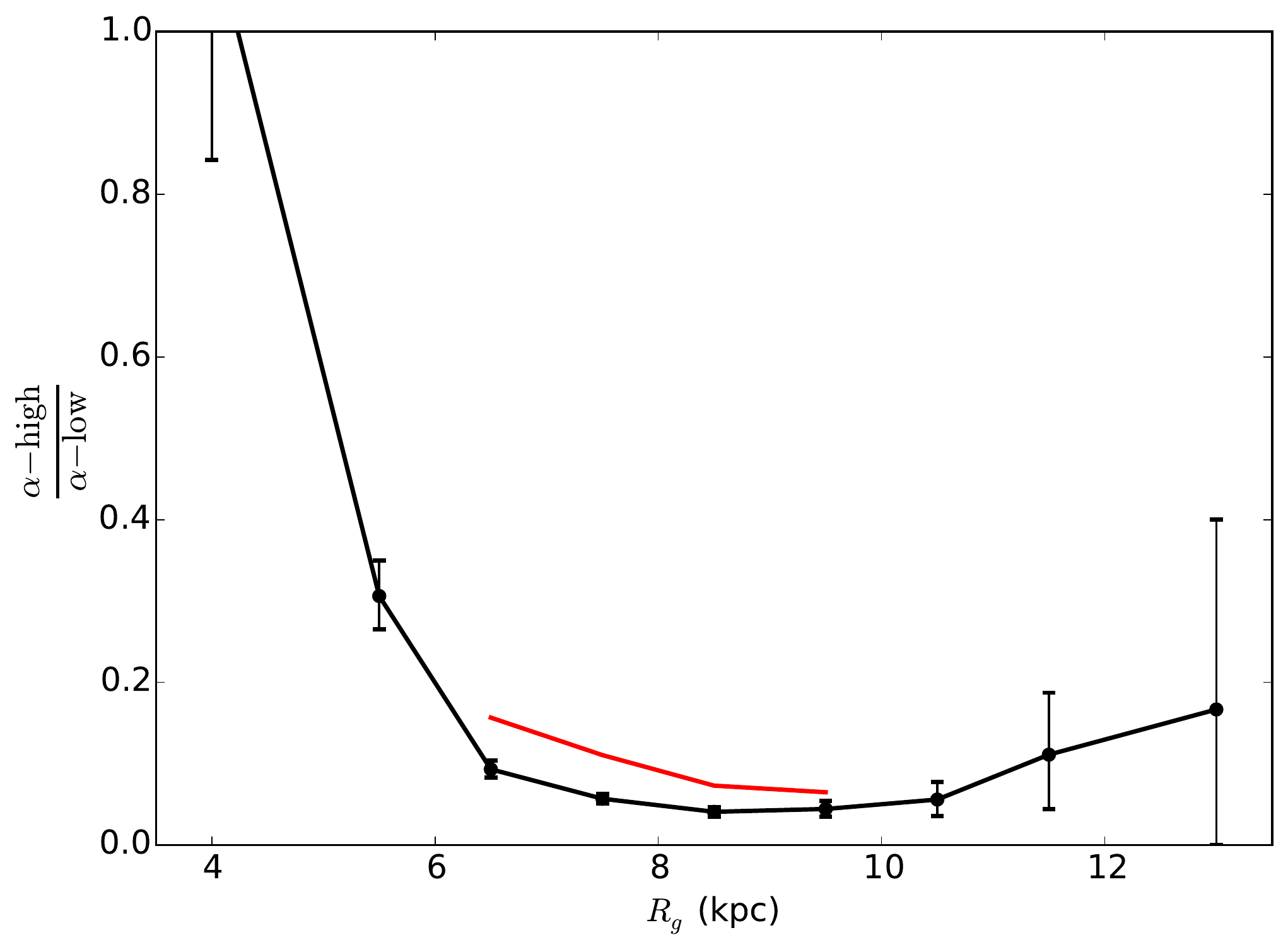}
\caption{Ratio of the  $\alpha$-high to  the $\alpha$-low populations (as defined by the red line of Fig.~\ref{fig:Rg_meta_bins}), for stars binned in guiding radii (black curve) or observed radii (red curve). The vertical error bars indicate Poisson uncertainties.}
\label{fig:Population_ratio}
\end{center}
\end{figure}

\bigskip
 We investigated the behaviour of the $\alpha$-high and
$\alpha$-low populations in different bins of guiding radii. The
two-dimensional histograms of $\aM$ vs $\meta$ as a function of $R_g$
are shown in Fig.~\ref{fig:alpha_meta_Rgbins}:
 overall, the same trends are recovered as those obtained by \citet[][see also \citealt{Nidever14}]{Anders14}, with however differences in details, described below.

The change in the  ratios between $\alpha$-high and $\alpha$-low populations is shown in Fig.~\ref{fig:Population_ratio} (black line).
The trend seen in Fig.~\ref{fig:Population_ratio} is consistent with a shorter radial scale-length for the $\alpha$-high population, as  proposed for example by \citet{Bovy12b} and \citet{Bensby14}, but our selection bias against inner-disc stars on a circular orbit prevents a more quantitative statement. 
 Interestingly the value of this ratio for stars with guiding centre radii close to the solar circle is  
$\sim
0.048$,  almost 50 per cent smaller than the ratio derived
when evaluating the ratio at a location based on  the observed position of the stars ($\sim 0.09$, shown in red in
Fig.~\ref{fig:Population_ratio}), the latter being consistent with the local  thin disc - thick disc ratio obtained by
other studies that defined the discs either geometrically \citep[e.g.][]{Juric08} or kinematically \citep[e.g.][]{Kordopatis13c}.

\smallskip

\begin{figure*}
\begin{center}
\includegraphics[width=\textwidth]{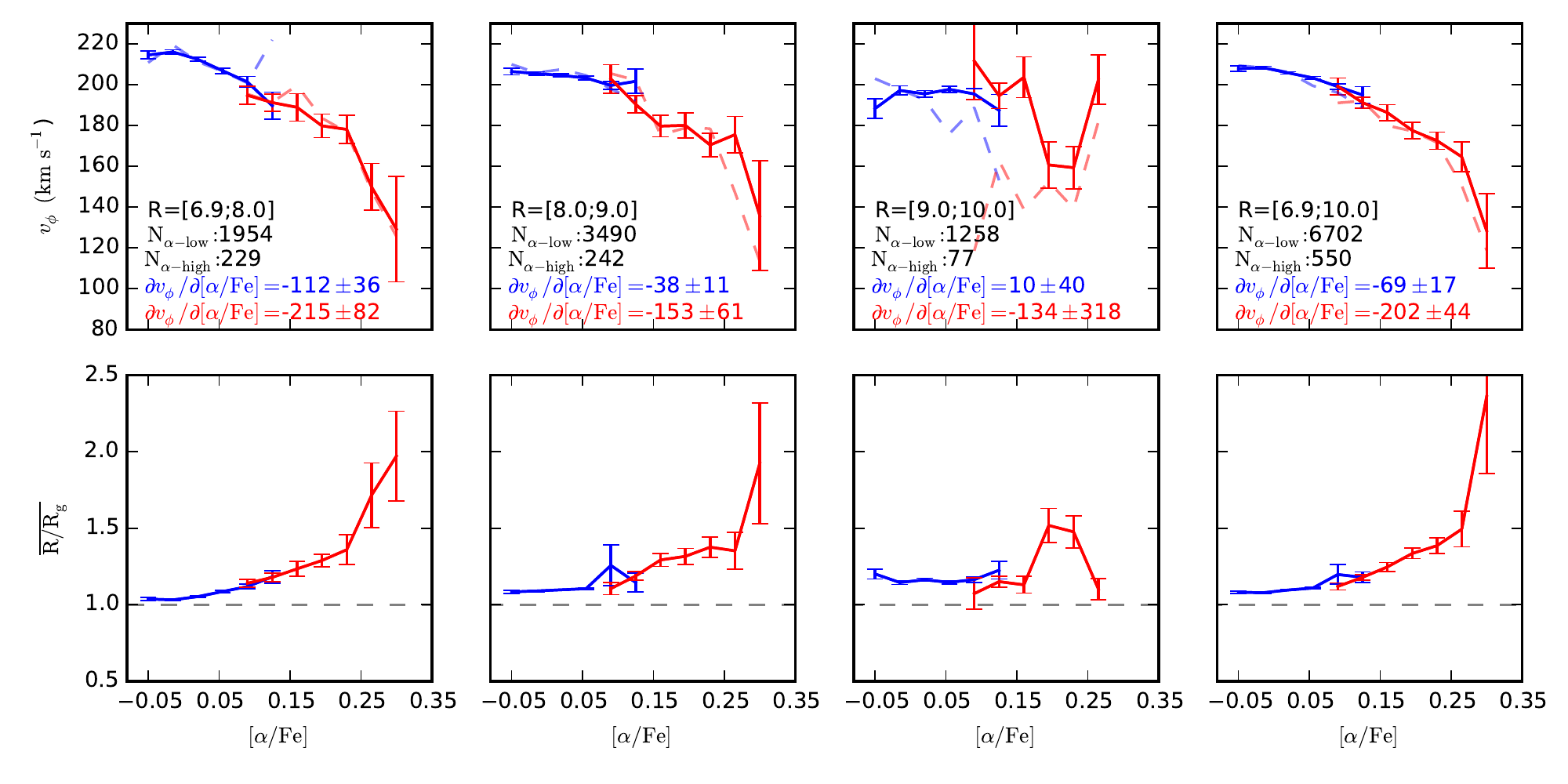}
\caption{Estimated trends of the mean rotational velocity (upper panels) and the ratio $R/R_g$ (lower panels) as a function of $\aM$ (bins of $0.035\dex$ overlapping by $0.01\dex$) for stars at different $R$ having $|Z|\leq0.5\kpc$.  The trends for the $\alpha$-high/thick disc and the $\alpha$-low/thin disc  are plotted in red and blue, respectively. Stars with negative $\vphib$ have been removed from the sample, and only those bins containing at least five targets are considered. The error bars are computed as $\sigma_x/\sqrt{N}$, where $x$ is the parameter under consideration and $N$  is the number of stars within a given bin. The solid  lines are obtained using Eq.~$\ref{eq:vrot}$, whereas the dashed lines (in the upper panels) show the trends using estimates of the azimuthal velocity that include proper motions.}
\label{fig:disc_trends_alpha}
\end{center}
\end{figure*}

\begin{figure*}
\begin{center}
\includegraphics[width=\textwidth]{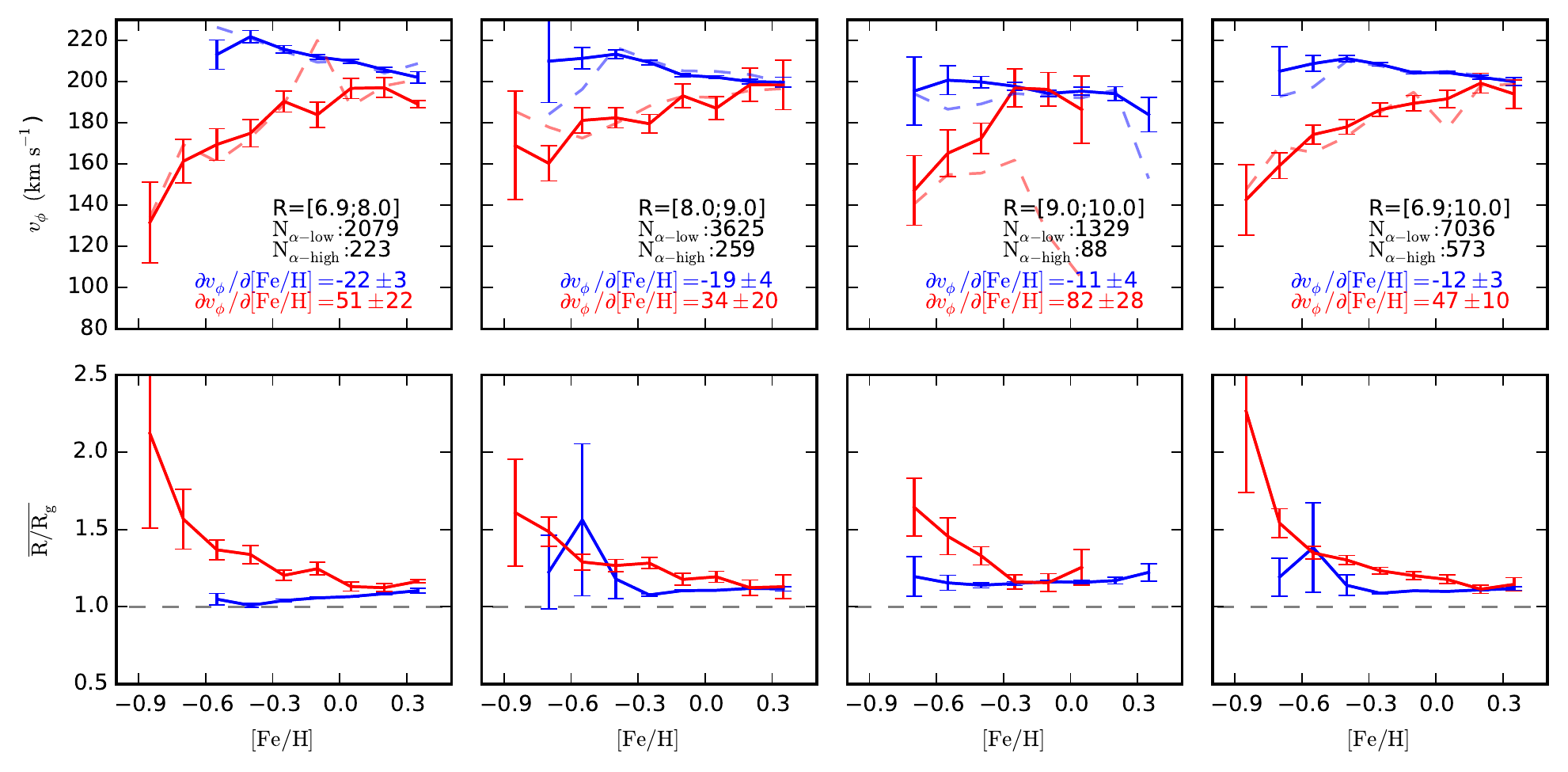}
\caption{Same as Fig.~\ref{fig:disc_trends_alpha} but showing  the iron abundance on the $x$-axis. The iron abundance bins are $0.15\dex$ wide and overlap by $0.05\dex$. }
\label{fig:disc_trends_feh}
\end{center}
\end{figure*}

We next compute the trends between the azimuthal velocity and the
chemical composition of the stars, using Eq.~\ref{eq:vrot} to estimate
the mean velocities in $0.15\dex$ width bins in $\meta$ and $0.035\dex$ width bins in $\aM$ (only the
bins that contain at least five stars are analysed).  We
consider all the stars closer than $0.5\kpc$ from the plane,  first divided in three radial bins of $1\kpc$ width, and second considering the entire available $R$ range. Furthermore, in order
to minimise contamination from the halo, we also exclude counter-rotating stars. 
The results are shown in
Fig.~\ref{fig:disc_trends_alpha} and Fig.~\ref{fig:disc_trends_feh}.

\begin{table*}
\caption{Correlation between $\vphib$ and the chemical composition of the stars}
\begin{center}
\begin{tabular}{|ccccc|}
\hline \hline
$R$ range & $(\partial \vphib / \partial \meta)_{\mathrm{\alpha-low}}$ & $(\partial \vphib / \partial \aM)_{\mathrm{\alpha-low}}$ &   $(\partial \vphib / \partial \meta)_{\mathrm{\alpha-high}}$ & $(\partial \vphib / \partial \aM)_{\mathrm{\alpha-high}}$ \\ 
$\kpc$		& $\kms\dex^{-1}$	& $\kms\dex^{-1}$	& $\kms\dex^{-1}$	& $\kms\dex^{-1}$ \\	\hline
$[6.9-8.0]$	& $-22\pm3$		& $-112\pm36$ 	& $51\pm22$ 	&$-215\pm82$\\ 
$[8.0-9.0]$ 	& $-19\pm4$		& $-38\pm11$ 		& $34\pm20$ 	&$-153\pm61$\\ 
$[9.0-10.0]$ 	& $-11\pm4$		& $10\pm40$ 		& $82\pm28$ 	&$-134\pm318$\\ \hline
$[6.9-10.0]$ 	& $-12\pm3$		& $-69\pm17$ 		& $47\pm10$ 	&$-202\pm44$\\ \hline
\end{tabular}
\end{center}
\label{tab:Correlations}
\end{table*}%

For the thin disc (identified as the $\alpha$-low population), we find the  $\partial \vphib / \partial x$ slopes, where $x$ is either $\afe$ or $\feh$,  both being negative, at all radii. The measured slopes (reported in columns 2 and 3  of Table~\ref{tab:Correlations}), are found to become flatter with increasing $R$.
It might be noted that the general decrease of $\aM$ with increasing $\meta$ seen for the $\alpha$-low sequence  leads to the expectation of opposite signs for these two gradients. That this is not what we find is possibly a manifestation of the changes in angular momentum as stars migrate, but needs to be confirmed by independent datasets.

Various values  for $\partial
\vphib / \partial \meta_{\mathrm{\alpha-low}}$ have been derived by others  (based on datasets of
different spectral resolution and surveyed volume) and range from
$-23$ to $-17\kms\dex^{-1}$, with typical reported errors of $4\kms\dex^{-1}$
\citep{Kordopatis11b,Lee11,Adibekyan13, Haywood13, Recio-Blanco14, Wojno16}. Our results are overall in agreement with these measurements.
To our knowledge, only one measurement of  $\partial \vphib / \partial \aM_{\mathrm{\alpha-low}}$ has been published in the literature, by \citet{Haywood13}, for a very local stellar sample ($d\lesssim 75$\,pc). Their value ($-3.1 \kms\dex^{-1}$) is lower than the one we derive at  $8<R<9\kpc$, though still in broad agreement (within $3\sigma$) with ours.

\bigskip 

Unlike the trends for the thin disc, the signs of the slopes of the $\vphib$ trends as a function of $\afe$ and $\feh$ are the same. Furthermore,  the mean velocity of the thick disc (identified as those stars in the $\alpha$-high population) does  not show an invariant  linear relation with $\meta$ but rather the slope flattens, and approaches that of the thin disc, at chemical abundances similar to those of the thin disc stars (Fig.~\ref{fig:disc_trends_feh}). 
This flattening is barely seen when using the $\vphipm$ values to calculate the mean kinematics; the trends for the two  $\alpha$-sequences in this case merge, as indicated by the dashed lines in the upper panels of Fig.~\ref{fig:disc_trends_feh}. We note, however, that given the errors in the proper motions highlighted in the analysis above, and  the improved accuracy of the line-of-sight estimators  $\vphib$, the flattening is likely to be a real feature of the data. Indeed, the previous reliance on azimuthal velocities derived using proper motions  may be the reason why this flattening remained unnoticed until now.

\smallskip

Nevertheless, it is not straightforward to ascertain  whether the trends towards decreasing values of $\partial \vphib / \partial \meta$ and $\partial \vphib / \partial \aM$ are: (i) intrinsic to the thick disc, (ii) related to contamination by thin disc stars in the high $\feh$ and low $\aM$ bins or (iii) are manifestations of the presence of   $\alpha$-high stars that have migrated outwards  from the inner disc \citep[e.g.][]{Masseron15,Chiappini15, Anders16}.

The dividing line between $\alpha$-high and $\alpha$-low sequences
 (Fig.~\ref{fig:alphafe_cardinal}) is less-well constrained at the high
 $\feh$ end, where the separation between the sequences is
 minimised, and this could lead to incorrect assignment of
 $\alpha$-low stars to the $\alpha$-high sequence. We emphasise,
 however, the fact that for iron abundances above $+0.2$ we find a
 lower value for $\vphib$ for the $\alpha$-high population than for
 the $\alpha$-low population. 
 This is indicative that the $\alpha$-high sequence that we defined does not consist purely of  `normal' thin disc stars (which would have larger values of $\vphib$) with large errors in their abundances (and hence wrongly assigned to the $\alpha$-high sequence).  Indeed, it could either be a signature of an age gradient as a function of metallicity along one of the sequences \citep[see for example][]{Haywood13},  combined with a  continuously varying asymmetric drift, and/or the extension of the thick disc up to these super-solar metallicity values. These statements are further supported by the higher value for the ratio  $R/R_g$ for thick disc (i.e.~$\alpha$-high) stars compared to that for thin disc (i.e.~$\alpha$-low) stars  at $\feh\gtrsim 0.2$ (see bottom plots of  Figs.~\ref{fig:disc_trends_alpha}, \ref{fig:disc_trends_feh} and Fig.~\ref{fig:APOGEE_Rg_vs_R}).

Fitting a first order polynomial to the trends for $\aM>0.1$ and $\meta<0$ for the thick disc yields the slopes reported in columns 4 and 5 of Table~\ref{tab:Correlations}. Unlike for the thin disc, no clear trend as a function of $R$ is seen. The correlations we derive are in good agreement (within one sigma) with earlier values from the literature derived from various datasets with different volume coverage \citep{Spagna10,Kordopatis11b,Lee11,Adibekyan13, Haywood13, Recio-Blanco14}.

\section{Constraints on the formation of the thick disc}
\label{sect:thick}

  The well-defined and steep relation in $\partial \vphib / \partial \meta$ that we find, places constraints on models of thick disc formation. Indeed, mixing, whether due to mergers or migration, act to lessen gradients and weaken such trends. 

 On the one hand, within a disc having a negative radial metallicity gradient, the `blurring' effect of epicyclic motions is to introduce a negative
$\partial \vphib / \partial \meta$ trend (whereby stars of different metallicities can have different azimuthal velocities due to the fact that they visit a given position while being close to the apocentre or pericentre of their orbit). On the other hand,  changes in the guiding radii of stars (i.e. churning) have been suggested by \citet{Loebman11} to add scatter,  smear out gradients  and eventually (in case of full mixing)  erase the $\partial \vphib / \partial \meta$ trend (see their Fig.~10 and associated discussion in their section 4.3).

On the basis of this insight, the steep trends that we
find for the thick disc (defined chemically as the $\alpha$-high
population) present a challenge for models in which radial migration plays a dominant
role.  Our finding
complements previous investigations, such as \citet{Kordopatis11b,
Minchev12c, Vera-Ciro14} or \citet{Aumer16} that conclude it is difficult to form a thick disc as observed 
through radial migration of stars from the inner regions of the disc. However, it should be noted that this does not demonstrate that 
radial migration has not taken place in the thick disc; indeed,
\citet{Solway12} have shown in their dynamical study that radial 
migration can affect stars on hot orbits, provided there exist transient spirals with the necessary open arm structure (for given spiral pattern, the decreasing effectiveness of radial migration is  evident in the negative correlation between rms change in angular momentum as a function of velocity dispersion, their Fig.~10).

\bigskip

 A possible explanation for the positive $\partial \vphi /
\partial \feh$ trend of the thick disc is given by
\citet{Schonrich16}, where the authors find such a trend in models with   
inside-out growth of the proto-thick disc. In their scenario, a very
rapid growth in the scale-length of the star forming gas in the early
Galaxy (on a comparable timescale to SNIa enrichment) leads to a positive $\partial \vphi / \partial \feh$
trend with amplitude comparable to our findings here.

 An alternative possibility to explain the positive trend that is being measured,  could be the rapid\footnote{Here, by rapid we mean a  couple of Gyr, to be compatible with the narrow age range of the bulk of thick disc stars and the lack of a significant metallicity gradient in the thick disc \citep{hayden14}.} collapse and spin-up  of a cloud  (via conservation of the angular momentum) within which most thick disc stars are born \citep{Jones83,Brook04}. Indeed, 
within any one self-enriching population, after the onset of
Type Ia supernovae there should be a steady decrease of [$\alpha$/Fe]
with time \citep[unless there are strong bursts of star formation;][]{Gilmore91}. 
Such a decline has been identified for thin disc \citep{Haywood13,Bensby14,Nissen15, Chiappini15} and thick disc
stars in the Solar vicinity  \citep[][]{Haywood13,Bensby14}.   Specifically, the thick disc exhibits a
strong correlation between $\alpha$-abundances and age: 
the most $\alpha$-enhanced thick disc stars are $\sim12-13\Gyr$ old,
and  the mean age drops to $\sim9\Gyr$ at the lowest values of
$\alpha$-enhancement\footnote{Note that \citet{Nissen15} did not find such a trend but his sample was limited to three thick disc stars.}. The very steep gradient  $\partial \vphib /\partial \aM$ that we find for the $\alpha$-high sequence could then indicate that this population increased its 
$\vphib$ within a short time, $\sim3\Gyr$,
  and also explain the positive $\partial \vphi / \partial \feh$ trend.

Though there are some possible hints from
Fig.~\ref{fig:Rg_meta_bins}, Fig.~\ref{fig:Rg_meta_bins_scatter} and Fig.~\ref{fig:alpha_meta_Rgbins}
 (see also
Fig.~\ref{fig:churning_attempt} in the next Section) that the
low-$\alpha$ and high-$\feh$ thick disc stars have larger guiding
radii than the thick disc stars of lower $\feh$ and higher
$\alpha$-abundance \citep[therefore pointing to inside-out growth characteristics, see][the second row of their Fig.~3, and \citealt{Schonrich16}]{Minchev14b}, 
any firm conclusion that such evidence has been
found is difficult to make due to small number statistics in our
sample and/or no optimal separation between $\alpha$-high and $\alpha$-low populations. 
Therefore, the radial collapse of a cloud cannot be excluded based on our analysis.

In particular, it is still unclear from the data whether the thick disc extends above solar metallicities or if the super-solar metallicity $\alpha$-high stars are part of a different population. Indeed, this chemical region has been found to host some surprisingly intermediate-age high-metallicity stars, that have been suggested by \citet{Martig16,Anders16} to be born at the end of the bar and then migrate to the solar radius. Although these peculiarly young stars are few and should not dominate our sample, the chemistry of the high-$\alpha$, high-$\feh$ stars is still compatible with stars that were born in the inner galaxy. The kinematics, close to the ones of the locally-born thin disc (nevertheless with slightly hotter velocities) is on the other hand also compatible with stars that have radially migrated via churning, while having also been mildly dynamically heated over time \citep[due to resonances at the outer Lindblad resonance with the bar, see][]{Debattista06,Minchev12c}.

\section{Putting constraints on the evolution of the thin disc and on churning}
\label{sect:constraints_churning}
 As already stated in the previous section, the $\alpha$-low stars exhibit only a weak correlation between their velocities and their chemical composition, result compatible with non-negligible effects of churning mechanisms. 
 This is in agreement with the large scatter in the
age-metallicity relation for the thin disc near the Sun  \citep[][]{Edvardsson93}.


\begin{figure}
\begin{center}
\includegraphics[width=0.5\textwidth]{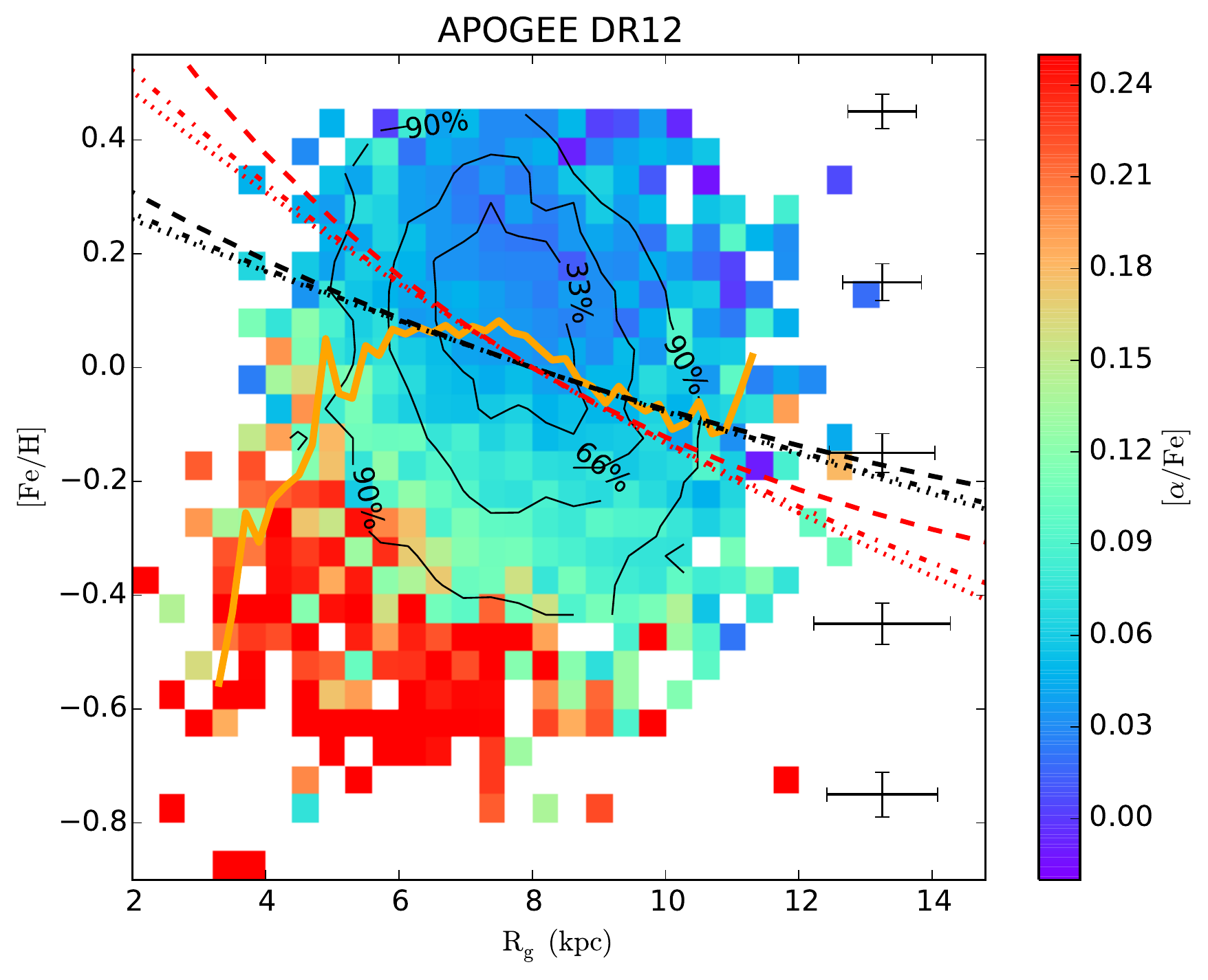}
\caption{Iron abundance of stars closer than $0.5\kpc$ from the
plane, as a function of guiding radius, colour-coded as a function of
the median $\afe$. This plot should be free of blurring
effects from epicyclic motions. The red and black dashed and dotted curves
indicate extreme examples of possible iron abundance gradient of the
ISM, following an exponential form (see
Eq.~\ref{eq:met_gradient}). Different values for $\feh$ at
large radii (controlled by the factor A) and for the local
ISM metallicity gradient have been adopted to illustrate possible
ranges for the birth radii of the stars.  The black
and red curves adopt, respectively, an ISM gradient of
$-0.04\dex\kpc^{-1}$ and $-0.07\dex\kpc^{-1}$. The orange solid curve represents the running mean of the iron abundance as a function of $R_g$ for this sample.}

\label{fig:churning_attempt}
\end{center}
\end{figure}

Figure~\ref{fig:churning_attempt} shows the iron abundance of the stars closer than $0.5\kpc$ to the Galactic plane plotted against their guiding radius, with bins   colour-coded as a function of the median $\aM$. Again, we recall that this plot is designed to reveal underlying trends between  chemistry and angular momentum,  after removal of the blurring effects of epicyclic motions.  As in \citet{Kordopatis15a}, the red and black lines illustrate the possible birth radii of  stars of a given iron abundance,   based on  two  choices for the local  $\feh$  gradients  in  the ISM, both assuming  the extreme case in which $\meta$ increases exponentially inwards with a scale length $R_M$, such that
\begin{equation}
 \meta(R)=A\cdot(1-e^{-(R-R_0)/R_M}).
 \label{eq:met_gradient}
 \end{equation}
 The constant $A$ in this formula is the value to which $\meta$ tends at large radii, and together with the value of local metallicity gas gradient,  sets  the  value  of $R_M$ ($R_M=A/{\rm ISM gradient}$).
The two estimates of the local ISM gradients that we adopt are $-0.04\dex\kpc^{-1}$ and $-0.07\dex\kpc^{-1}$, respectively in black and red, that include the range of gradients found in the literature, based on different metallicity tracers \citep[e.g.][]{Boeche14,Genovali14}. We show curves for three  values of $A$, namely $-0.5, -1.0, -1.5$.

Given the spatial limitations of our cardinal-direction sample  (no stars with $R<6.9\kpc$), we expect that the distribution  in Fig.~\ref{fig:churning_attempt} is biased against small $R_g$ values. In particular,  stars in circular orbits with  small $R_g$ should not be present in our sample. Due to the age-velocity dispersion and age-metallicity relations resulting in different asymmetric drifts for the stellar populations, plus the radial gradient in thin-disc velocity dispersion, our sample is biased against typically thin disc stars in the inner Galaxy (expected to be  preferentially metal-rich). Such a lack of metal-rich stars at small $R_g$ is reflected in the derived trend of the $\meta$ gradient (orange line in Fig.~\ref{fig:churning_attempt}): for $R_g\lesssim 6.5\kpc$, the  metallicity gradient of our sample does not follow any of the plausible behaviours described by Eq.~\ref{eq:met_gradient}, and indeed the metallicity decreases with decreasing radius. 

In contrast,  our selection should not introduce a significant metallicity bias against the outer Galaxy sample. Further, 
according to \citet{Hayden15}, the selection function of APOGEE should also not influence  the final results much, provided that a relatively small range in $Z$-height is included \citep[see also][for details concerning the APOGEE selection function]{Zasowski13}.
The stars that we select towards the cardinal directions, with $|\beta| \geq 0.8$, satisfy this vertical restriction,  as can be seen from the $Z$ distribution of Fig.~\ref{fig:Galactic_positions}. It then follows that 
 the stars that lie above the red (and/or black) lines  in Fig.~\ref{fig:churning_attempt} are unlikely to have been born at the present value of  their guiding radius, but rather have had their orbital angular momentum increased by disc asymmetries.

 In particular, we find stars with $\meta \sim 0.4$ and $R_g\sim 10\kpc$, and $\meta>0.2$ and $R_g>11\kpc$ (where $\meta_{\rm ISM}\approx -0.1$).  As one would expect from radial migration  (in combination with a mean metallicity increasing with time), the maximal $\feh$ value reached by the super-metallicity stars at a given $R_g$ is a decreasing function of $R_g$. 
 Indeed, depending on the temporal and spatial properties of the structures/gravitational perturbations in the disc,  churning can in principle affect most of the stars in the disc, 
being more effective for stars that are kinematically cold.  The stars in the innermost parts of the Galaxy  (where the ISM has the highest iron abundance) are kinematically hotter, and  whereas a star could in principle be caught in corotation resonances of  successive spiral arms  over a wide range of $R$, the probability of this happening declines with the required distance to migrate  \citep[e.g.][]{Schonrich09a,Minchev12c,Daniel15}. As a consequence of the two points above,  not only should the maximal $\feh$ reached at a given $R_g$ decline, but also one should expect  a positive  age gradient as a function of $R_g$ {\bf ($\partial Age / \partial R_g$)}  for stars having iron abundance higher than that of their local ISM.

Deriving ages from spectroscopically measured atmospheric parameters
is challenging, with relative errors typically at least  30 per cent
\citep[e.g.][]{Jorgensen05, Soderblom10}. However, for red giant stars the surface abundance ratio of C/N is modified by hot-bottom-burning and subsequent dredge-up and thus this abundance ratio, in combination with stellar evolutionary models,  can be used to infer stellar mass and age 
\citep{Masseron15, Martig16,Ness16}. Calibration with stars with asteroseismology data from the Kepler satellite allowed \citet[][see their Fig.~5]{Martig16} to conclude that for metallicities above Solar, [C/N]$\lesssim -0.45$ corresponds to
ages $\sim 2\Gyr$, whereas [C/N]$\gtrsim -0.1$ corresponds to ages
$\gtrsim 6\Gyr$.

We adopted the derived ages of \citet{Martig16} to investigate how the stellar ages change as a function of $R_g$ for the thin disc, restricting the sample to only those stars with $|Z|\leq0.5\kpc$. 
 
\begin{figure}
\begin{center}
\includegraphics[width=0.5\textwidth]{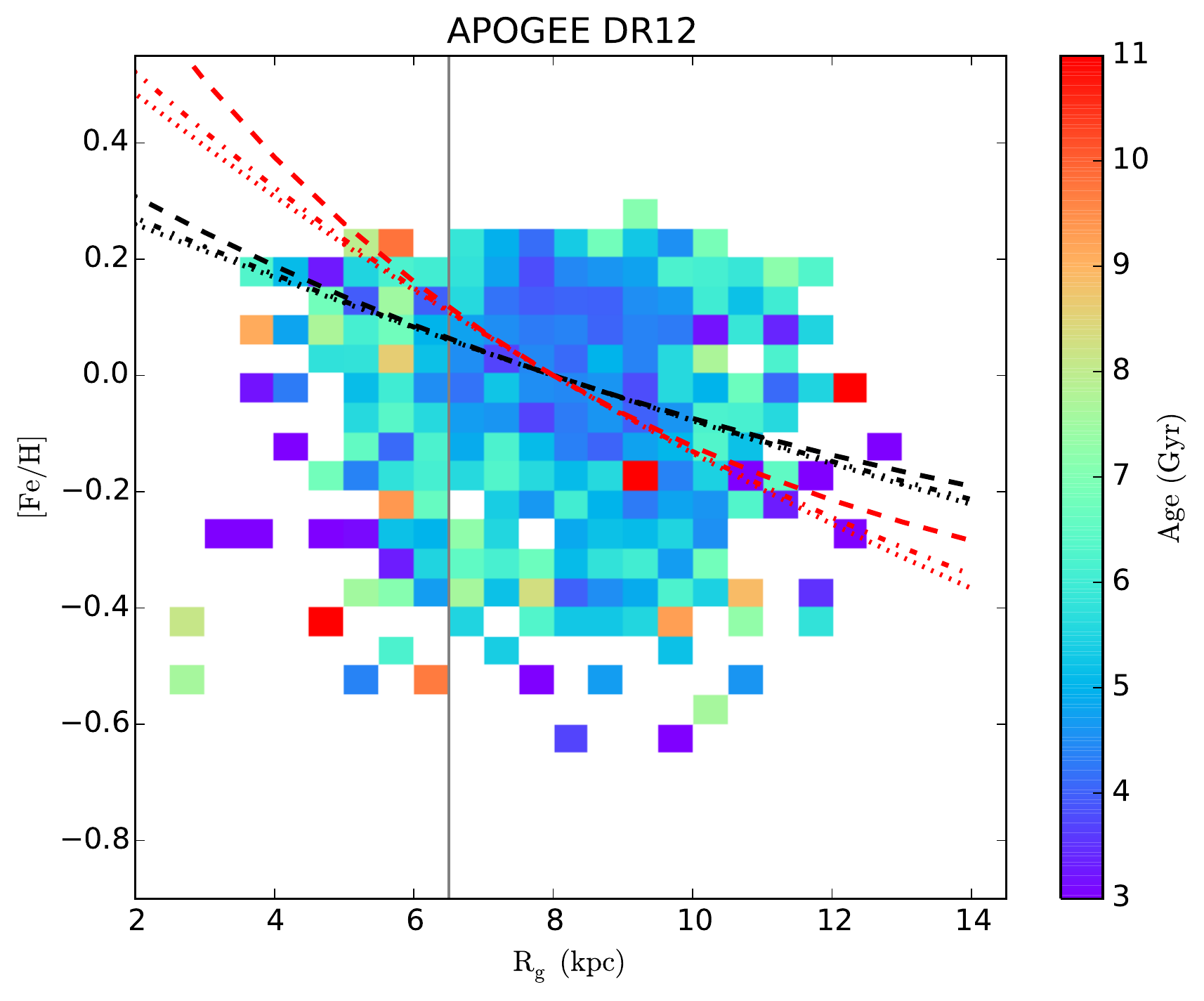}
\caption{As Fig.~\ref{fig:churning_attempt}, but only for thin disc stars and colour-coding now representing the median ages, as derived by \citet{Martig16}. Fewer  targets  are represented than in Fig.~\ref{fig:churning_attempt}, due to either the unavailability of $C$ and $N$ measurements or the  non-validity of the calibration relation used by \citet{Martig16} to obtain the ages. The grey vertical line indicates the $R_g$ below which our sample is biased against metal-rich stars.}
\label{fig:churning_attempt_ages}
\end{center}
\end{figure}

The median ages as a function of $R_g$ and \meta\, are illustrated in Fig.~\ref{fig:churning_attempt_ages}. 
We find that  for a given range of iron abundance or guiding radius there is a wide spread of ages. Furthermore, in agreement with \citet{Chiappini15}, we find a metal-poor ($\feh\lesssim -0.3$) young ($\lesssim 4\Gyr$) population in the inner disc, which given our sample selection (no stars with $R\lesssim6.5\kpc$) should be on high eccentricity orbits in order to enter our volume.

In addition,  there is a hint for older high-metallicity stars to be located in the outer regions. This is consistent with radial migration of stars from the inner parts of a disc that formed inside-out: stars require time in order to get caught in successive co-rotation resonances and the time needed is a function of the distance travelled away from the stellar birth-place. 
However, the selection function of APOGEE against young stars and the limited spatial range of our sample prevents us from drawing further conclusions without fully taking into account these selection effects. 
In particular, it would be of great interest to measure variations of mean age as a function of $R_g$ for different metallicity bins. 

 Negative age gradients are typical signatures of inside-out growth of the disc: the inner parts of the disc form stars more rapidly, and enrich faster, that the outer regions  \citep[e.g.][]{Matteucci89, Pagel97, Chiappini97, Roskar08, Schonrich16}. Stars of a given iron abundance that do not migrate will therefore be younger in the outer disk than in the inner disc. One could hence investigate the way $\partial Age /\partial R_g$ changes as a function of metallicity to constrain the relative importance of radial migration, as opposed to conditions at birth, on the chemo-dynamical structure of the thin disc. Future data, with minimal biases in  age and spatial distribution, in combination with better ages coming either from astero-seismology or Gaia, will shed further light and allow the quantification of this effect.

\section{Conclusions}
\label{sect:conclusions}

We have investigated the increased precision with which one component of space motion can be measured by restricting a sample of stars to only those in lines-of-sight close to a Cardinal direction. We demonstrated that a restriction to stars lying towards the direction of
Galactic rotation allow us to probe the azimuthal velocities with a  precision better than $\sim 20\kms$. We  have also demonstrated that caution needs to be taken when analysing extreme velocities derived using proper motions.
 Application of this los technique to the APOGEE survey defines, with unprecedented precision, the correlations
between $\vphib$, $\meta$ and $\aM$ for the $\alpha$-high and
$\alpha$-low stars, usually identified as the chemical thick and thin
discs, respectively. The trends that we measured confirmed that these
two populations are not only chemically different, but also exhibit
different kinematics, likely resulting from the different dynamical processes that were important during their evolution.

Combination of the azimuthal velocities with distances allow the orbital angular momenta to be derived, and from these the guiding centre radii,  with a precision of $\sim1\kpc$. Use of the guiding radii as opposed to the observed radii eliminates the blurring effects of epicyclic motions in the  spatial and chemical distributions of the stars.
Our results confirm that $\alpha$-high stars are more radially concentrated than the  $\alpha$-low disc stars.  
Furthermore, we identified for the thin disc clear  signatures of radial migration as well as a flattening of the $\partial \vphib / \partial \afe$ and $\partial \vphib / \partial \feh$ with Galactocentric radius. We  excluded  radial migration from being important in the formation of the chemically defined thick disc, on the basis of a very strong correlation between the azimuthal velocity and the $\aM$ abundances (in agreement with previous analyses).  

   Future papers of this series will analyse samples selected from the APOGEE survey 
in the other cardinal directions, probing the vertical and radial
motions of the stars, and for the Gaia-ESO and RAVE surveys. We expect to gain  valuable insight into the chemo-dynamic evolution of the Milky Way Galaxy.

\section*{Acknowledgments}
We thank the anonymous referee for comments and suggestions that improved the quality of the manuscript.  We also thank M.\,Martig, M.~Hayden, J.\,Read and C. Scannapieco for useful  discussions about the transverse velocities, ages and proper motion errors. M. Maia and N. da Costa are 
 acknowledged regarding the stellar distances used in this paper. 
RFGW thanks the Leverhulme Trust for support through a Visiting Professorship at the University of Edinburgh.

\appendix
\section{Error on the measured mean velocities}
\label{app:error_mean_velocity}
The error on the measured  mean velocity $\overline{v}$ of a population $N$ stars, having an intrinsic velocity dispersion $\sigma_v$  is given by:
\begin{equation}
\label{eqn:error_mean}
error(\overline{v})=\sqrt{\frac{\sigma_m^2 + \sigma_v^2}{N}},
\end{equation}
where   $\sigma_m$ is the typical uncertainty on an  individual measurement of velocity.

Consider two samples, $A$ and $B$, with samples sizes of
$N_A=100$ and $N_B=15$ and typical errors $\sigma_{m,A}\approx60\kms$
and $\sigma_{m,B}\approx15\kms$. These two cases are designed to mimic those of  stars in a chemical abundance bin
of a wide-area survey using space motions derived from ground-based proper
motions, such as previous analyses of the Gaia-ESO survey
\citep[e.g.][]{Kordopatis15b} (sample A) and those appropriate for a sample restricted in lines-of-sight and using a lower-uncertainty  velocity estimator, such as the present paper (sample B).  

Adopting  $\sigma_v=5\kms$ (as appropriate for a cold stream or substructure), this gives:
\begin{eqnarray}
error(\overline{v_A})&=&6\kms \\
error(\overline{v_B})&=&4.1\kms.
\end{eqnarray}
For  $\sigma_v=20\kms$ (consistent with estimates of the azimuthal velocity dispersion in old thin-disk stars), then:
\begin{eqnarray}
error(\overline{v_A})&=&6.3\kms \\
error(\overline{v_B})&=&6.4\kms.
\end{eqnarray}

Despite the reduced sample size, sample B, with lower uncertainties, does as well, if not better, than the larger sample with higher uncertainties, in terms of mean trends.


\section{Validation of the azimuthal velocity estimations with red clump stars}
\label{sec:rc_validation}
\begin{figure*}
\begin{center}
\includegraphics[width=\textwidth]{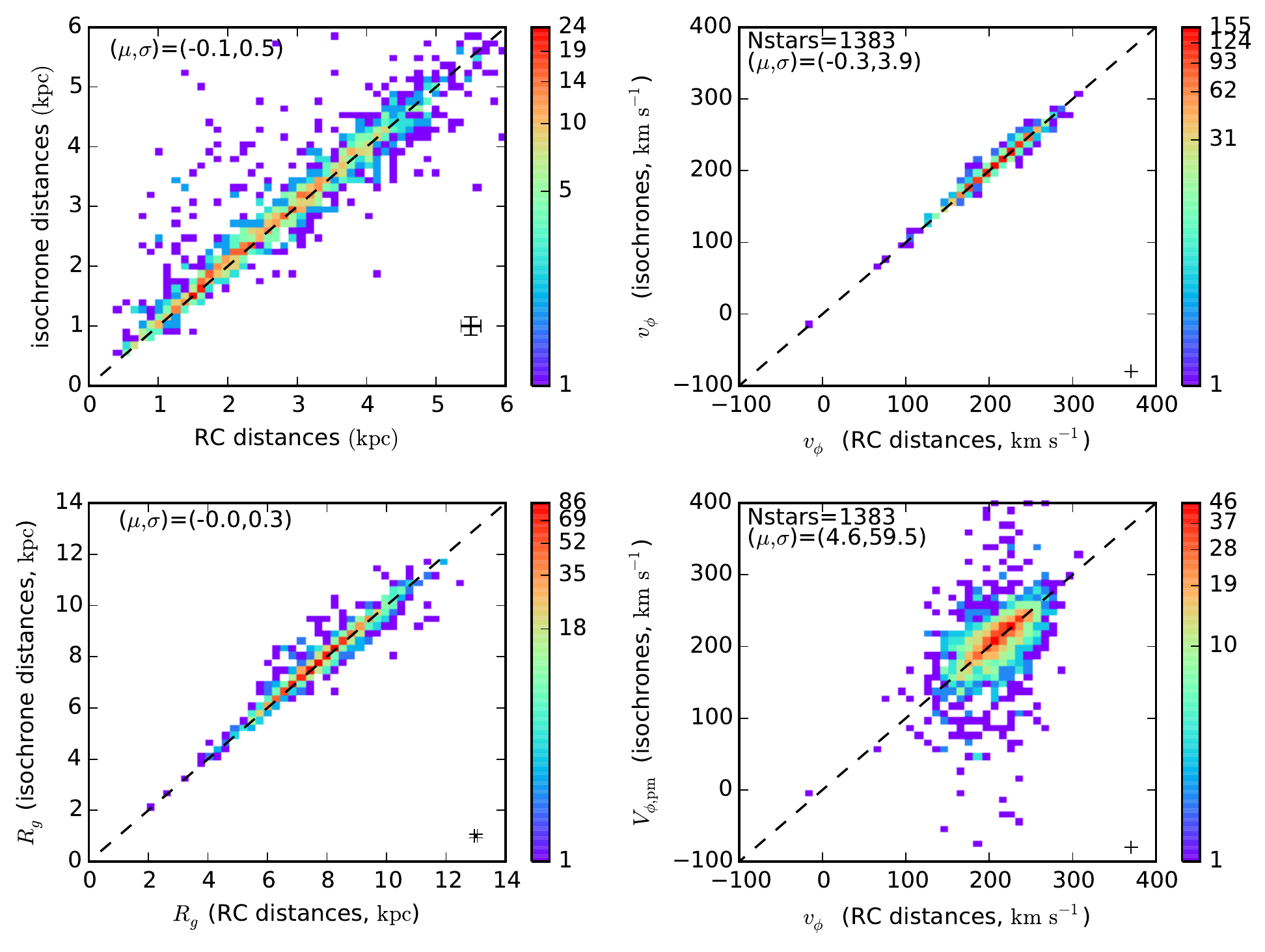}
\caption{ Upper left: Comparison of the distances obtained via isochrone fitting \citep[$y$-axis,][]{Santiago16} versus those derived from the distance modulus with an adopted  red-clump luminosity \citep[$x$-axis,][]{Bovy14a}. Upper right: Comparison of the $\vphib$ velocities (obtained without the proper motions) computed using each of these two distance estimates. Lower left: Comparison between the guiding radii obtained using $\vphib$ and each of the two distance estimates. Lower right: Comparison between the azimuthal velocities obtained using the proper motions in combination with the isochrone distances ($y$-axis) and the azimuthal velocities obtained without proper motions and with red-clump distances ($x$-axis). The sample size together with the values of the mean difference and the dispersions between the two parameters are indicated in the upper left of each panel. }
\label{fig:RC_comparisons}
\end{center}
\end{figure*}

Here, we investigate using the APOGEE DR12 red clump (RC) subsample towards the cardinal direction , the effect of uncertainty on the stellar distances. 
Distances for RC stars are precise to  5 per cent and unbiased to 2 per cent \citep{Bovy14a}. Out of the 6925 stars in the cardinal directions, 1383 are RC stars.
The line-of-sight distances  obtained with an isochrone fitting technique or based on the distance modulus of the RC (right plot of Fig.~\ref{fig:RC_comparisons}) are in good overall agreement, however,  several kiloparsecs are noted between the two distance estimates for a non-negligible amount of stars. 

In the upper right plot of Fig.~\ref{fig:RC_comparisons}, are compared the \emph{estimated} $\vphib$ (ie obtained without proper motions) when the two distance estimates are used. It can be seen that the differences in the distances do not affect much the final velocity estimation (since distance is multiplied by trigonometric functions when computing $\vphib$).
In fact, for the 1383 stars the bias is only $-0.3\kms$ and the dispersion is $3.9\kms$. 

The Lower right plot compares the two most extreme cases of the azimuthal velocity estimations. On the $y$ axis is the least precise measurement that one can have using the isochrone distances and the proper motions, and on the $x$ axis are the $\vphib$ estimates using the RC distances (\ie the most precise measurement). As expected, one can see that the discrepancies are significant, illustrating the need of our estimate of $\vphib$ to derive accurate chemodynamical properties of the Galaxy.

Finally, the lower left plot shows how the guiding radii compare, when using the RC distances and the isochrone distances. One can see that the agreement is overall good. 

The above plots indicate, firstly, that uncertainties in distances are not the ones driving the uncertainties in $\vphib$. Indeed, the dominant factor is the uncertainty in the proper motions. Furthermore, these plots suggest that we can use the entire APOGEE sample towards the cardinal directions, without limiting ourselves to the RC, which is biased against metal-poor stars and contains a factor of five fewer stars.

\bibliographystyle{mnras}
\def\aj{AJ}\def\apj{ApJ}\def\apjl{ApJL}\def\araa{ARA\&A}\def\apss{Ap\&SS}
\def\mnras{MNRAS}\def\aap{A\&A}\def\nat{Nature}
\def\nar{New Astron. Rev.}

\bibliography{../No_pm}
\end{document}